\documentclass[prd,twocolumn,showpacs,nofootinbib,amsmath,amssymb]{revtex4-1}
\usepackage{graphicx,dcolumn,bm}
\usepackage{hyperref}
\usepackage{amssymb}
\usepackage{color}
\begin{document}

\newcommand{\MRF}[1]{{\bf \color{red}{[MRF: #1]}}}
\newcommand{\SA}[1]{{\bf \color{blue}{[SA: #1]}}}
\newcommand{\MF}[1]{{\bf \color{green}{[MF: #1]}}}

\title{How bright can the brightest neutrino source be?}

\author{Shin'ichiro Ando}
\affiliation{GRAPPA Institute, University of Amsterdam, 1098 XH
Amsterdam, The Netherlands}
\author{Michael R. Feyereisen}
\affiliation{GRAPPA Institute, University of Amsterdam, 1098 XH
Amsterdam, The Netherlands}
\author{Mattia Fornasa}
\affiliation{GRAPPA Institute, University of Amsterdam, 1098 XH
Amsterdam, The Netherlands}
\date{January 9, 2017; revised \today}
\begin{abstract}
After the discovery of extraterrestrial high-energy neutrinos, the next
 major goal of neutrino telescopes will be identifying astrophysical
 objects that produce them.
The flux of the brightest source $F_{\rm max}$, however, cannot be
 probed by studying the diffuse neutrino intensity.
We aim at constraining $F_{\rm max}$ by adopting a broken power-law flux
 distribution, a hypothesis supported by observed properties of any
 generic astrophysical sources.
The first estimate of $F_{\rm max}$ comes from the fact that we can only
 observe one universe, and hence, the expected number of sources above
 $F_{\rm max}$ cannot be too small compared with one.
For abundant source classes such as starburst galaxies, this one-source
 constraint yields a value of $F_{\rm max}$ that is an order of
 magnitude lower than the current upper limits from point-source searches.
Then we derive upper limits on $F_{\rm max}$ assuming that the
 angular power spectrum is consistent with neutrino shot noise yet.
We find that the limits obtained with upgoing muon neutrinos in IceCube
 can already be quite competitive, especially for rare but bright source
 populations such as blazars.
The limits will improve nearly quadratically with exposure, and
 therefore be even more powerful for the next generation of neutrino
 telescopes.
 \end{abstract}
\pacs{95.85.Ry, 98.70.Vc}
\maketitle

\section{Introduction}
\label{sec:intro}

IceCube firmly detected astrophysical neutrinos, but currently, it is not
possible to identify a neutrino source and the distribution of neutrino
events is consistent with being isotropic~\cite{IceCubeScience,
Aartsen:2014gkd, Aartsen:2015knd, IceCubePS, IceCubeAnis, Aartsen:2016xlq}.
Accumulating more and more data of the diffuse intensity will sharpen
constraints on an {\it average} source flux, but the flux of the
{\it brightest} source cannot be probed directly with this approach as
long as the distribution remains consistent with isotropic.
How bright can the brightest neutrino source be?
This is the next question that needs to be addressed.
Searches for point-like sources determined that the upper limit (post-trial
and per neutrino flavor) on the flux of the brightest neutrino source,
$F_{\rm max}$, ranges from $2 \times 10^{-12}$ to $3 \times
10^{-11}$~TeV~cm$^{-2}$~s$^{-1}$, depending on declination $\delta$ and
assuming $E^{-2}$ energy spectrum~\cite{IceCubePS}.

Here, we address the same question by taking a different approach.
In particular, we implement a {\it statistical distribution} for the
flux of neutrino sources, a more realistic hypothesis than the
single-flux population assumed in, e.g., Refs.~\cite{IceCubePS,
IceCubeAnis}.
By constraining the shape of the source flux distribution with
observables such as the intensity and anisotropies of the diffuse 
neutrinos, we will derive constraints on $F_{\rm max}$.

Our approach is twofold. First, we discuss estimates on $F_{\rm max}$ that
are intrinsic to the fact that we only have access to one universe to
sample the source distribution.
If the expected number of sources at $F_{\rm max}$ becomes much
smaller than one, then it is unlikely that one could observe larger
fluxes in this universe.
We show that if the number of sources producing the diffuse neutrino
flux measured by IceCube is greater than $\sim$10$^3$, then this
one-source limit of $F_{\rm max}$ is smaller than the upper limits from
Ref.~\cite{IceCubePS}.
Thus, our findings allow us to make statements for a flux regime that is
still unprobed by IceCube.

Recent analysis of the angular power spectrum found no significant
clustering of multiple events~\cite{IceCubeAnis}.
As our second approach, we set upper limits on $F_{\rm max}$ based on
this null result, and show that they are tighter than what is inferred
from the search for point-like sources, at least for rare source
populations.
These constraints on $F_{\rm max}$ are effective in a regime where the
one-source limit is above the point-like source limit, showing that the
two strategies followed are complementary.
We find that the method is particularly constraining
even with the current IceCube exposure if we adopt upgoing muon neutrino
events~\cite{Aartsen:2016xlq}, which would provide a critical test for
blazar interpretation as the origin of the diffuse neutrino flux.
We also find that the limits obtained from the angular power spectrum
improve quadratically with the exposure.
Thus, they provide an extremely powerful probe for the next generation
of neutrino telescopes, such as IceCube-Gen2~\cite{IceCubeGen2} and
KM3NeT~\cite{KM3NeT}.


In this paper, we constrain the flux of the brightest source (rather
than, e.g., its joint luminosity and distance), as it is the quantity
that is directly relevant to detectability of the neutrino sources---a
goal yet to be achieved.
Although the flux is a phenomenological quantity, this way,
we can make our discussions model independent.
Another complementary approach would be to use typical luminosity and
density of each source.
Although these are more physical quantities, the discussions tend to be
highly model dependent.
We provide useful conversion formulae for a representative case.

This paper is organized as follows.
After introducing relevant formulation of the flux distribution and its
relation to the intensity and angular power spectrum in
Sec.~\ref{sec:Formalism}, we discuss current constraints on $F_{\rm
max}$ using the one-source argument and the angular power spectrum in
Secs.~\ref{sec:One-source constraint} and \ref{sec:Angular power
spectrum}, respectively. In Sec.~\ref{sec:Application to known source
populations}, we apply these generic discussions to several cases of known
source populations. Section~\ref{sec:Prospects for the future} is 
then devoted to what is expected in the future, before briefly concluding in 
Sec.~\ref{sec:Conclusions}.

\section{Formulation}
\label{sec:Formalism}

\begin{figure}
 \begin{center}
  \includegraphics[width=8.5cm]{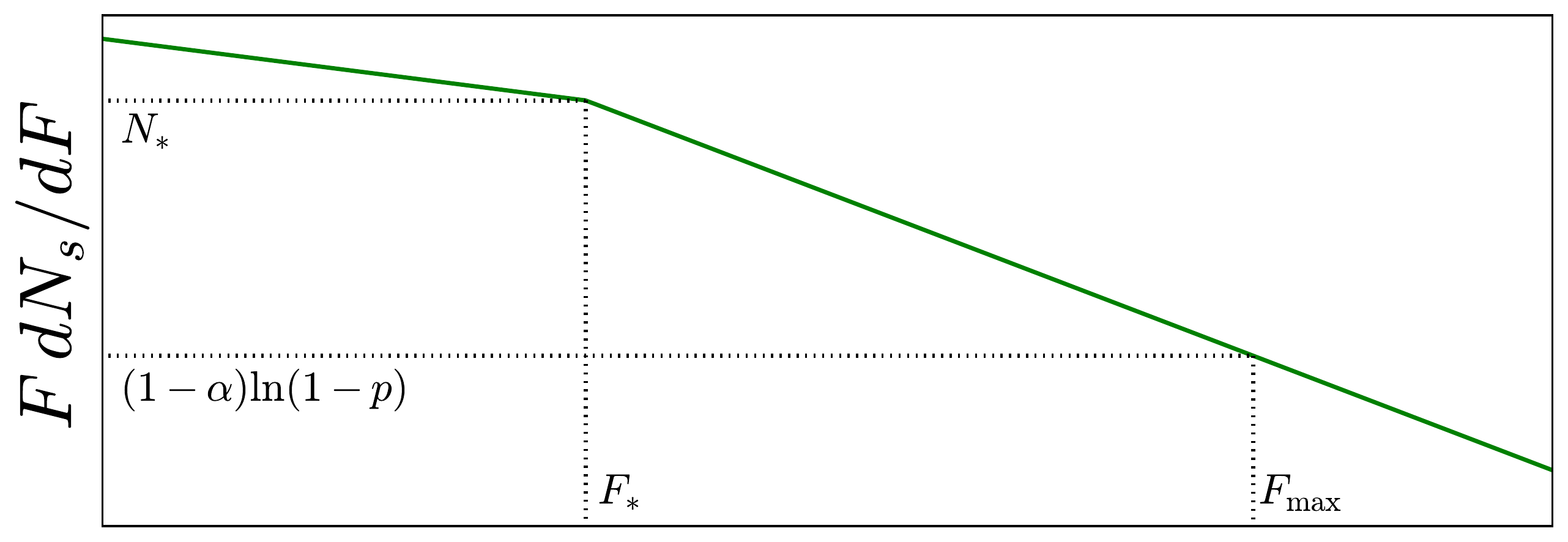}
  \includegraphics[width=8.5cm]{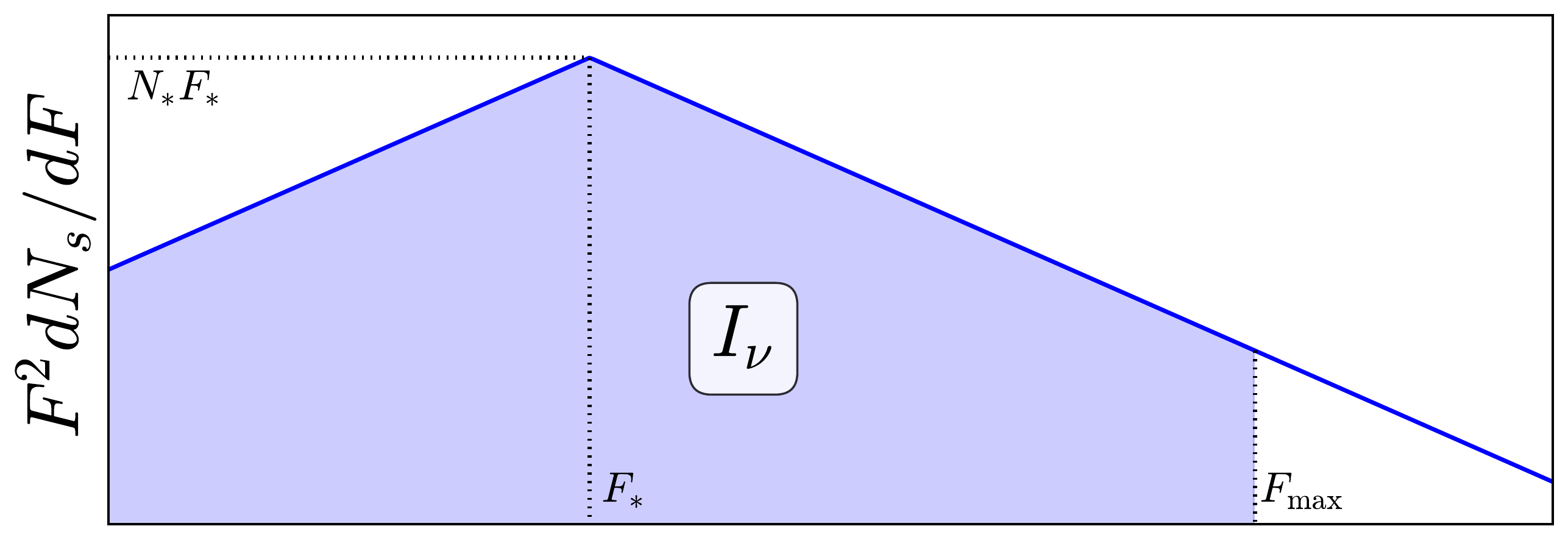}
  \includegraphics[width=8.5cm]{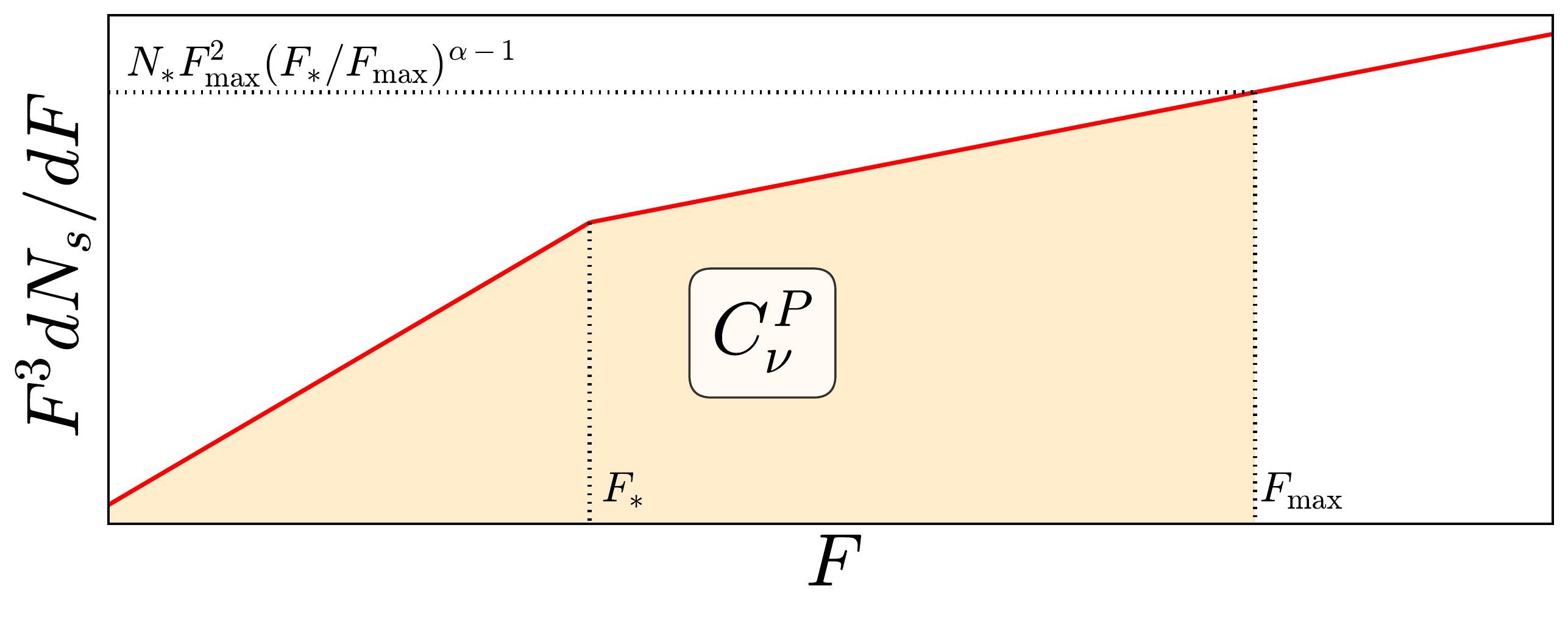}
  \caption{The source flux distribution $dN_s/dF$ multiplied by $F$ (top), $F^2$
  (middle), and $F^3$ (bottom), for $2<\alpha <3$ and $1<\beta
  <2$. Both horizontal and vertical axes are in logarithmic scales. The
  shaded regions in the middle and bottom panels represent that areas
  below these broken lines correspond to the intensity $I_\nu$
  [Eq.~(\ref{eq:nuintensity})] and the Poisson angular power spectrum
  $C_\nu^P$ [Eq.~(\ref{eq:Poisson angular power spectrum})],
  respectively; i.e., $I_\nu$ and $C_\nu^P$ are dominated by sources
  near $F_\ast$ and $F_{\rm max}$, respectively.}
  \label{fig:dNdF}
 \end{center}
\end{figure}

We define $N_s$ as the total number of sources from all sky and 
$\mathcal N_s = N_s / 4\pi$ as their surface number density.
The source flux distribution function is defined as $dN_s/dF$ and we
also use the equivalent probability density function of the single source 
$P_1(F) \equiv d\ln N_s/dF$.
Our hypotheses on the form of $P_1(F)$ are rather mild: We assume that
the distribution follows a broken power-law with physically
motivated parameters. In particular, $\alpha$ denotes the slope of the 
distribution, $P_1(F) \propto F^{-\alpha}$, above a 
characteristic flux $F_\ast$. We assume $2<\alpha<3$, which is compatible with
what is observed in sources detected in other wavelengths such as gamma
rays, e.g., blazars~\cite{BLLacLF,FSRQLF,Ajello:2015mfa,Inoue:2015kuy},
star-forming galaxies~\cite{TAM, Feyereisen2016}, and radio
galaxies~\cite{DiMauro:2013xta,Hooper2016}.
In fact, if these sources are distributed homogeneously in a
local volume where cosmological effects can be ignored ($z\ll 1$), it is
well known that the flux distribution reduces to the Euclidean limit,
i.e., $\propto F^{-5/2}$~\cite{Peacock}.
This is expected, in particular, for the brightest sources (since these
are likely to be nearer to us than the fainter members of their source
class), and therefore, $\alpha = 2.5$ will be our reference value.
For fluxes smaller than $F_\ast$, the slope of the distribution must
flatten in order to avoid divergences (cf. Olbers' paradox). We assume
$P_1(F) \propto F^{-\beta}$ for $F<F_\ast$ with $\beta < 2$. The
flattening of the slope at low fluxes is, again, supported
observationally~\cite{BLLacLF,FSRQLF,Ajello:2015mfa}.
The top panel of Fig.~\ref{fig:dNdF} schematically shows this
distribution.
A discussion of flux distributions with the assumption $\alpha<2$ on the
power-law slopes is postponed until Appendix~\ref{app:Case of a flat
distribution}.

In a pixel with a size $\Omega_{\rm pix}$ that roughly corresponds to
the angular resolution of the detector, there are on average $N_s^{\rm
pix}$ sources, with $N_s^{\rm pix} = \mathcal N_s \Omega_{\rm pix}$.
Then, the flux per pixel is given by the sum of the fluxes of $N_s^{\rm
pix}$ individual sources.\footnote{In general, $N_s^{\rm pix}$ is non-integer, and thus a more precise expression is given by a convolution
with a Poisson distribution.}
The mean and variance of the flux distribution {\it per pixel}, $P(F)$,
is simply given by $N_s^{\rm pix}$ times the mean and variance of the
flux distribution {\it per source}, $P_1(F)$:
\begin{eqnarray}
 \langle F \rangle &=& N_s^{\rm pix} \langle F \rangle_{P_1}, \\
  \langle (F-\langle F\rangle)^2\rangle &=& N_s^{\rm pix}\langle
   (F-\langle F\rangle_{P_1})^2\rangle_{P_1},
\end{eqnarray}
where $\langle \cdot \rangle$ and $\langle \cdot \rangle_{P_1}$ indicate
averages taken over $P(F)$ and $P_1(F)$, respectively. Under our
assumptions for $P_1(F)$, it is straightforward to show that
\begin{eqnarray}
 \langle F\rangle_{P_1} &\simeq& \eta_1 F_\ast^2 P_1(F_\ast), \\
  \langle (F-\langle F\rangle_{P_1})^2\rangle_{P_1} &\simeq& \langle F^2
   \rangle_{P_1}   \nonumber \\ &=& \eta_2 F_{\rm max}^3
   P_1(F_{\rm max}),
   \label{eqn:variance} 
\end{eqnarray}
where $\eta_1 = (\alpha-2)^{-1} + (2-\beta)^{-1}$ and $\eta_2 =
(3-\alpha)^{-1}$ are both constants of order unity.
Note that, in Eq.~(\ref{eqn:variance}), instead of integrating up to
infinity, we truncated at $F_{\rm max}$.
We define $N_\ast^{\rm pix}$ as the typical number of sources 
per pixel around flux $F_\ast$, i.e., $N_\ast^{\rm pix} \equiv N_s^{\rm
pix} F_\ast P_1(F_\ast)$, and similarly, we define $N_\ast$ and
$\mathcal N_\ast$ corresponding to $N_s$ and $\mathcal N_s$,
respectively.
Then, we obtain the following for the first two moments of the flux
distribution:
\begin{eqnarray}
 \langle F \rangle &=& \eta_1 N_\ast^{\rm pix} F_\ast, \\
 \langle (F - \langle F\rangle)^2\rangle &=& \eta_2 N_\ast^{\rm pix} F_{\rm max}^2
  \left(\frac{F_\ast}{F_{\rm max}}\right)^{\alpha-1}.
\end{eqnarray}
Equivalently, the intensity $I_\nu$ of the neutrino flux (also often
referred to as $\phi_\nu$) and its Poisson angular power spectrum
$C_\nu^P$ are, respectively,
\begin{eqnarray}
 I_\nu &=& \mathcal \eta_1 \mathcal N_* F_*, \label{eq:nuintensity}\\
 C_\nu^P &=& \eta_2\mathcal N_* F_{\rm max}^2
  \left(\frac{F_*}{F_{\rm max}}\right)^{\alpha-1}.
  \label{eq:Poisson angular power spectrum}
\end{eqnarray}
The middle and bottom panels of Fig.~\ref{fig:dNdF} show the flux
distribution multiplied by appropriate powers of $F$ such that the area
below the curves is proportional to $I_\nu$ and of $C_\nu^P$,
respectively.

In the following, expressions with an explicit index $E$, such as
$I_\nu(E)$ and $C_\nu(E)$, represent differential quantities with
respect to energy, and those without the index are the quantities
integrated over the energy.

\section{One-source constraint}
\label{sec:One-source constraint}

We are limited to observe a single universe, which then limits our
capability to constrain physical quantities.
Specifically, we cannot probe arbitrarily large fluxes, because
once the number of sources expected at such fluxes becomes 
smaller than one, it is unlikely to reconstruct the distribution in
the region.
We define the one-source limit on the flux of the brightest neutrino
source, $F_{\rm max}^{\rm 1s}$, such that only with a small probability
$p$ could we find at least one source brighter than $F_{\rm max}$ in the
entire sky.


The mean number of sources above $F_{\rm max}$ is given by $N_s
\Psi_1(>F_{\rm max})$, where $\Psi_1 (>F_{\rm max})$ is the
complementary cumulative distribution function corresponding to
$P_1(F)$.
Using the Poisson distribution with this mean, the probability $1-p$
of finding no source brighter than $F_{\rm max}$ is $\exp[-N_s
\Psi_1(>F_{\rm max})]$.
By solving this for a power-law $P_1(F) \propto F^{-\alpha}$, we obtain
\begin{equation}
 F_{\rm max} P_1(F_{\rm max}) = \frac{1-\alpha}{N_s}\ln (1-p),
  \label{eq:cosmicvariancelimit}
\end{equation}
which further translates into
\begin{equation}
 F_{\rm max} = \frac{I_\nu}{\eta_1\mathcal{N}_*}
 \left[\frac{4\pi\mathcal{N}_*}{(1-\alpha)\ln (1-p)}\right]^{1/(\alpha-1)}.
  \label{eq:cosmicvariancelimit2}
\end{equation}
In Eq.~(\ref{eq:cosmicvariancelimit}), $F_{\rm max}$ depends 
only on the properties of the source distribution function.
In Eq.~(\ref{eq:cosmicvariancelimit2}), on the other hand, it is recast
in terms of the measured intensity $I_\nu$ and the free parameter
$\mathcal{N}_*$.
For the Euclidean case ($\alpha = 2.5$), $F_{\rm max} \propto
I_\nu \mathcal N_*^{-1/3}$.
We assume that the intensity refers to neutrinos {\it per flavor}, and
where necessary, that flavor democracy holds, i.e., $I_{\nu_e} =
I_{\nu_\mu} = I_{\nu_\tau}$.
For an assumed $E^{-2}$ energy spectrum (in order to allow a direct
comparison with earlier results~\cite{IceCubePS}), $E^{2} I_\nu(E) =
(0.84 \pm 0.3) \times 10^{-11}$ TeV~cm$^{-2}$~s$^{-1}$~sr$^{-1}$, even
though a softer spectrum $E^{-2.58}$ provides a better
fit~\cite{IceCubeICRC}.

 \begin{figure}
 \begin{center}
  \includegraphics[width=8.5cm]{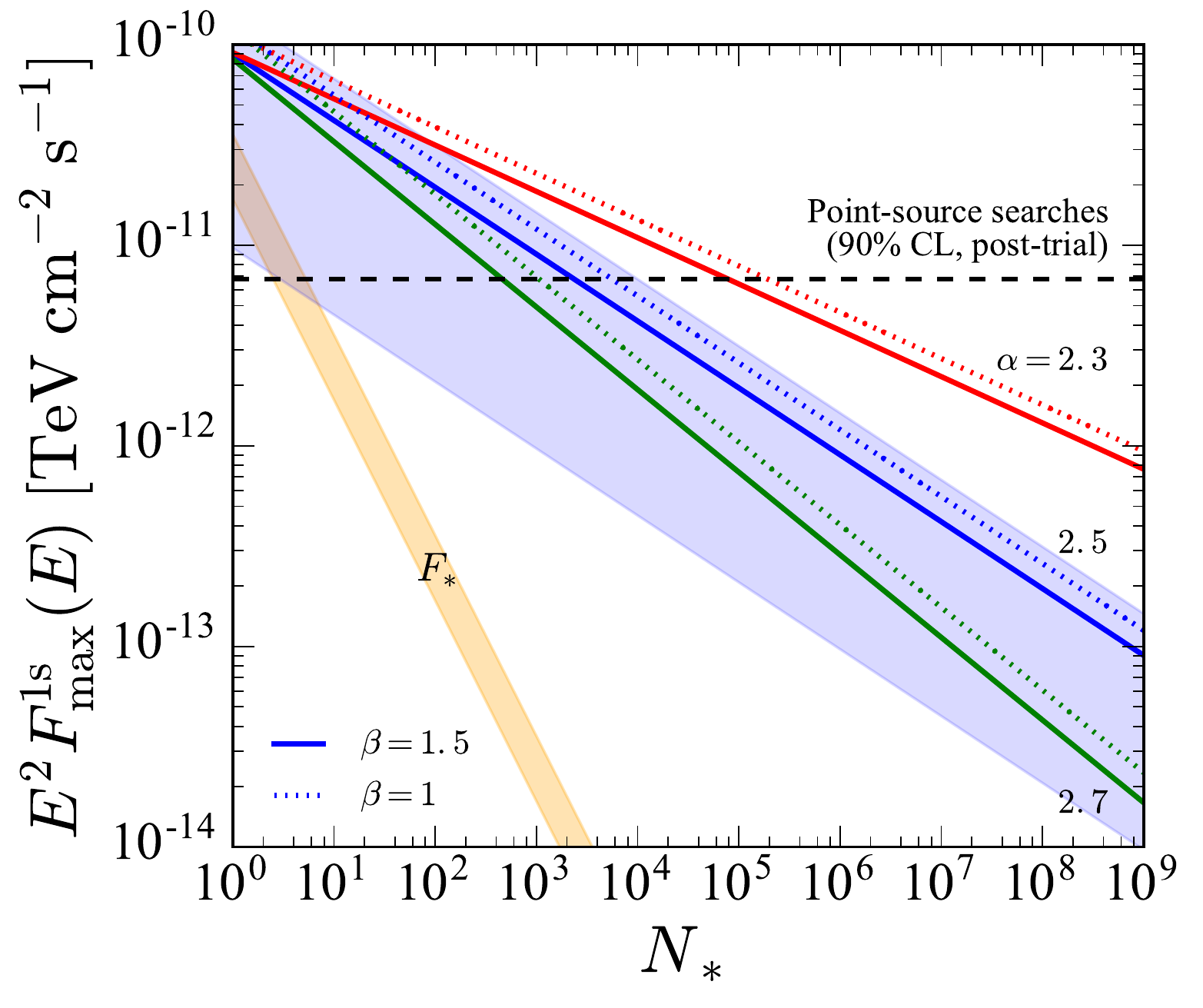}
  \caption{One-source upper limits (90\% CL) on the neutrino flux per
  flavor from the brightest neutrino source, as a function of the
  characteristic source number $N_*$, for various values
  of $\alpha$ and $\beta$. $F_{\rm max}^{\rm 1s}$ is defined from 
  Eq.~(\ref{eq:cosmicvariancelimit2}) as the flux for which there is a 90\% 
  probability of not finding any brighter source (solid and dotted). The
  blue band represents the region where the brightest source is located
  at 90\% CL for given $N_\ast$, in the Euclidean case with
  $(\alpha,\beta) = (2.5, 1.5)$. The
  dashed horizontal line represents the upper limit from the search for
  point-like source in Ref.~\cite{IceCubePS} toward the South Pole (see
  also Appendix~\ref{app:PS}). The orange band shows the characteristic
  flux $F_\ast$ of a single source required for the population from
  which it is drawn to explain the observed intensity $I_\nu$ according
  to Eq.~(\ref{eq:nuintensity}).}
  \label{fig:Fmax_CV}
 \end{center}
\end{figure}

Figure~\ref{fig:Fmax_CV} shows the one-source limits on the flux of the brightest source, $F_{\rm
max}^{\rm 1s}$, as a function of $N_\ast$ obtained with
Eq.~(\ref{eq:cosmicvariancelimit2}) for a few values of $\alpha$ and
$\beta$. For ease of comparison with the existing literature, these upper limits are presented at 90\% confidence
level (CL; $p = 0.1$).\footnote{Taylor expanding $F^\mathrm{1s}_\mathrm{max}$ for small $p$,
the reader may approximately rescale these upper limits from a significance $p^{(1)}$ to any desired significance $p^{(2)}$ with the ratio
$F^{(2)} =
\left[p^{(1)}/p^{(2)}\right]^{1/(\alpha-1)}\left[1+(p^{(1)}-p^{(2)})/(2(\alpha-1))\right]
F^{(1)}$.
The upper limit clearly gets weaker when $p^{(2)} < p^{(1)}$.
}
For $\alpha =2.5$ and $\beta = 1.5$, Eq.~(\ref{eq:cosmicvariancelimit2})
yields  $E^2 F_{\rm max}^{\rm 1s}(E) = 9.0 \times
10^{-11}~\mathrm{TeV~cm^{-2}~s^{-1}} / N_\ast^{1/3}$.
For comparison, we also show $F_\ast$ from Eq.~(\ref{eq:nuintensity})
with its uncertainty from the estimated error on $I_\nu$ (orange band),
and the upper limit from the search for point-like
sources~\cite{IceCubePS} (horizontal dashed line).
For derivation of the latter, see Appendix~\ref{app:PS}; see also
Ref.~\cite{Ahlers:2014ioa} for an estimate of the sensitivity when the
source density is modeled to follow the star-formation rate.

For source numbers $N_\ast$ greater than around $\sim$10$^3$, the
one-source limits reach below the upper limit from the search for
point-like sources~\cite{IceCubePS}.
In other words, finding a source at the flux level close to the
point-source upper limits for a source population characterized with
$N_* \gg 10^3$ (and $\alpha = 2.5$ and $\beta = 1.5$) is unlikely
with a chance probability of $p \approx 0.0016 (N_*/10^7)^{-1/2}$.

The flux cutoff is caused by either an intrinsic cutoff of the luminosity
function or by the volume effect, the latter of which is the case for
Euclidean sources ($\alpha = 2.5$; see Appendix~\ref{sec:Relation to
source density and luminosity}).
Then, Eq.~(\ref{eq:cosmicvariancelimit2}) can be regarded as a
prediction of $F_{\rm max}$.
For a given $N_\ast$, $F_{\rm max}$ has to be located between the values
of Eq.~(\ref{eq:cosmicvariancelimit2}) evaluated with $p = 0.05$ and $p
= 0.95$, at 90\% CL.
This is shown as a blue band in Fig.~\ref{fig:Fmax_CV} for $(\alpha,
\beta) = (2.5, 1.5)$.

We note that it is possible for the modeled population of sources to
give only a subdominant contribution to the diffuse neutrino intensity.
Indeed, Refs.~\cite{Bechtol:2015uqb,Glusenkamp:2015jca, Murase:2015xka}
suggest that neither starbursts nor blazars can explain the entirety of
the observed neutrino flux.
In that case, the one-source constraints become even tighter, as
$I_{\nu}$ in Eq.~(\ref{eq:cosmicvariancelimit2}) should be
replaced by $k I_\nu$, where $k$ is the fraction of the measured
intensity explained by the source class under investigation.
Having $k<1$ in Eq.~(\ref{eq:cosmicvariancelimit2}) will improve these
limits considerably.

\section{Angular power spectrum}
\label{sec:Angular power spectrum}

The maximum flux $F_{\rm max}$ can also be constrained by measuring the
variance of the source flux distribution; this information is essentially equivalent to the angular power spectrum. Indeed, if $F_{\rm max}$ is
too large, only a few of the brightest sources would be enough to make
the distribution of neutrinos highly anisotropic by yielding clustered
events, in conflict with what is measured~\cite{IceCubeAnis}.

\subsection{Formalism}

The number of neutrino counts per pixel $N_\nu^{\rm pix}$ is obtained by
multiplying the flux per pixel by the exposure, i.e., the product of the
effective area and the live time of the telescope.
Note that since the energy spectra of the
astrophysical and atmospheric neutrinos differ, so do the corresponding
exposures for each component, denoted by $\mathcal E$ and $\mathcal
E_{\rm atm}$, respectively.
The probability distribution of the number of
neutrinos per pixel
$N_\nu^{\rm pix}$ is therefore obtained by convolving the per-pixel flux distribution $P(F)$ and the Poisson
distribution with mean $F\mathcal E+F_{\rm atm} \mathcal E_{\rm atm}$:
\begin{equation}
P(N_\nu^{\rm pix}) = \int \mathcal P\left(N_\nu^{\rm pix}|F\mathcal E+F_{\rm
 atm}\mathcal E_{\rm atm}\right) P(F) dF,
\end{equation}
where $F_{\rm atm}$ is the flux of the atmospheric backgrounds, which
are assumed to be isotropic.
It is straightforward to obtain the moments of the distribution of
$N_\nu^{\rm pix}$:
\begin{eqnarray}
 \langle N_\nu^{\rm pix}\rangle
  &=& \langle F\rangle \mathcal E +F_{\rm atm}\mathcal E_{\rm atm},\\
 \langle (N_\nu^{\rm pix} - \langle N_\nu^{\rm pix}\rangle )^2\rangle &=&
  \langle  (F - \langle F\rangle )^2\rangle \mathcal E^2 + \langle
  N_\nu^{\rm pix}\rangle.
  \label{eq:count variance}
\end{eqnarray}
The first term of Eq.~(\ref{eq:count variance})
corresponds to the Poisson angular power spectrum that originates from
discreteness of the sources $C_\nu^P$ [Eq.~(\ref{eq:Poisson angular
power spectrum})], and the second corresponds to the shot-noise of the
neutrinos,
\begin{equation}
 C_\nu^N \equiv \frac{I_\nu}{\mathcal E} + \frac{\mathcal N_{\rm
  atm}}{\mathcal E^2},
\end{equation}
where $\mathcal N_{\rm atm} \equiv F_{\rm atm}\mathcal E_{\rm atm}
/ \Omega_{\rm pix}$ is the surface density of atmospheric background
events (see, e.g., Refs~\cite{AK, AKNT, Fornasa2016} in the case
of gamma rays).

The rms error for the angular power spectrum at multipole $\ell$ is
\begin{equation}
 \delta C_\ell = \sqrt{\frac{2}{(2\ell+1)f_{\rm sky}}}
  \left(C_\nu^{P} + \frac{C_\nu^N}{W_\ell^2}\right),
  \label{eq:deltaCl}
\end{equation}
where $f_{\rm sky}$ is a fractional sky coverage and $W_\ell$ is a beam
window function corresponding to the angular resolution of
IceCube~\cite{AK, AKNT, Fornasa2016}.
Since the purpose of this study is to obtain a simple estimate of the
current limits and future sensitivity rather than accurate values,
we assume $W_\ell = \exp(-\ell^2 \theta_{\rm
psf}^2/2)$. Given the null results from the anisotropy analysis~\cite{IceCubeAnis},
we estimate the upper limits on the Poisson angular power spectrum with
\begin{equation}
 C_\nu^P < \sigma \left(\sum_{\ell}\frac{1}{\delta
		   C_\ell^2}\right)^{-1/2},
 \label{eqn:upper_limit_APS}
\end{equation}
where $\sigma = 1.28$ (1.64) corresponds to the limits
at 90\% (95\%) CL.
By solving this as an equality for $C_\nu^P$, we obtain $C_{\nu,{\rm lim}}^P$ such that $C_\nu^P < C_{\nu,{\rm lim}}^P$. Then, by using Eqs.~(\ref{eq:nuintensity}) and (\ref{eq:Poisson angular
power spectrum}), we obtain the corresponding upper limits on $F_{\rm max}$ as
\begin{equation}
 F_{\rm max}^{\rm APS} <
  \frac{I_\nu}{\mathcal N_\ast}
  \left(\frac{\eta_1^{\alpha-1}}{\eta_2} \frac{\mathcal
   N_\ast C_{\nu,{\rm lim}}^P}{I_\nu^2}\right)^{1/(3-\alpha)}.
  \label{eq:Fmax_APS}
\end{equation}

To summarize, our estimates of $F^\mathrm{APS}_\mathrm{max}$ will rely
on observable inputs ($I_\nu$, $\mathcal N_{\rm atm}$), instrumental
inputs $(\mathcal{E}, f_\mathrm{sky}, \theta_{\rm psf})$, and theoretical inputs $(\alpha, \beta, N_\ast)$, which we will discuss for different source populations in Sec.~\ref{sec:Application to known source populations}. We present this analysis applied to two of the ``clean'' datasets of high-energy neutrinos from IceCube.

\subsection{High-Energy Starting Events (HESE)}

Since we care about the angular power spectrum of astrophysical sources,
we consider in the first instance only the High-Energy Starting Events
(HESE) dataset~\cite{IceCubeICRC}, a relatively clean event sample
consisting of showers and contained tracks at the highest energies.

We estimate $C_\nu^N = N_\nu / (4\pi \mathcal E^2)$ by using $N_\nu =
14$ (39) and four years of IceCube exposure for the muon (electron and
tau) neutrinos for the tracks (showers), a full-sky coverage $f_\mathrm{sky}=1$, the energy-dependent HESE effective area (from 1~TeV to 10~PeV) from
Ref.~\cite{IceCubeScience}, and the live time of the telescope (taken
accordingly to be 1347 days). The expected number of neutrinos is consistent with the
results of the four-year searches from Ref.~\cite{IceCubeICRC}: For an
energy spectrum proportional to $E^{-2}$, we find the total number of
neutrinos $4\pi I_\nu \mathcal E  = 26.1$. The rest of the measured events should be attributed to atmospheric 
backgrounds and statistical fluctuations. We also adopt angular resolutions of the order of the median angular resolution of the HESE events, namely $\theta_{\rm psf} =
1^\circ$ and 20$^\circ$ for tracks and showers respectively.

With these parameters, we obtain an upper limit on the Poisson angular power spectrum of
\begin{equation}
E^4 C_{\nu,{\rm lim}}^P(E) = 1.7\times 10^{-23} ~
\mathrm{TeV^2~cm^{-4}~s^{-2}~sr^{-1}},
\end{equation}
for the HESE tracks and
\begin{equation}
E^4 C_{\nu,{\rm lim}}^P(E) = 7.5 \times 10^{-22} ~ \mathrm{TeV^2~cm^{-4}~s^{-2}~sr^{-1}},
\end{equation}
for the HESE showers. Since the track events provide tighter constraints by more than one
order of magnitude, in the following, we will focus only on the flux limits
due to the tracks, so the intensity $I_\nu$ used in Eq.~(\ref{eq:Fmax_APS}) is that
of the muon flavor.

 \begin{figure}
 \begin{center}
  \includegraphics[width=8.5cm]{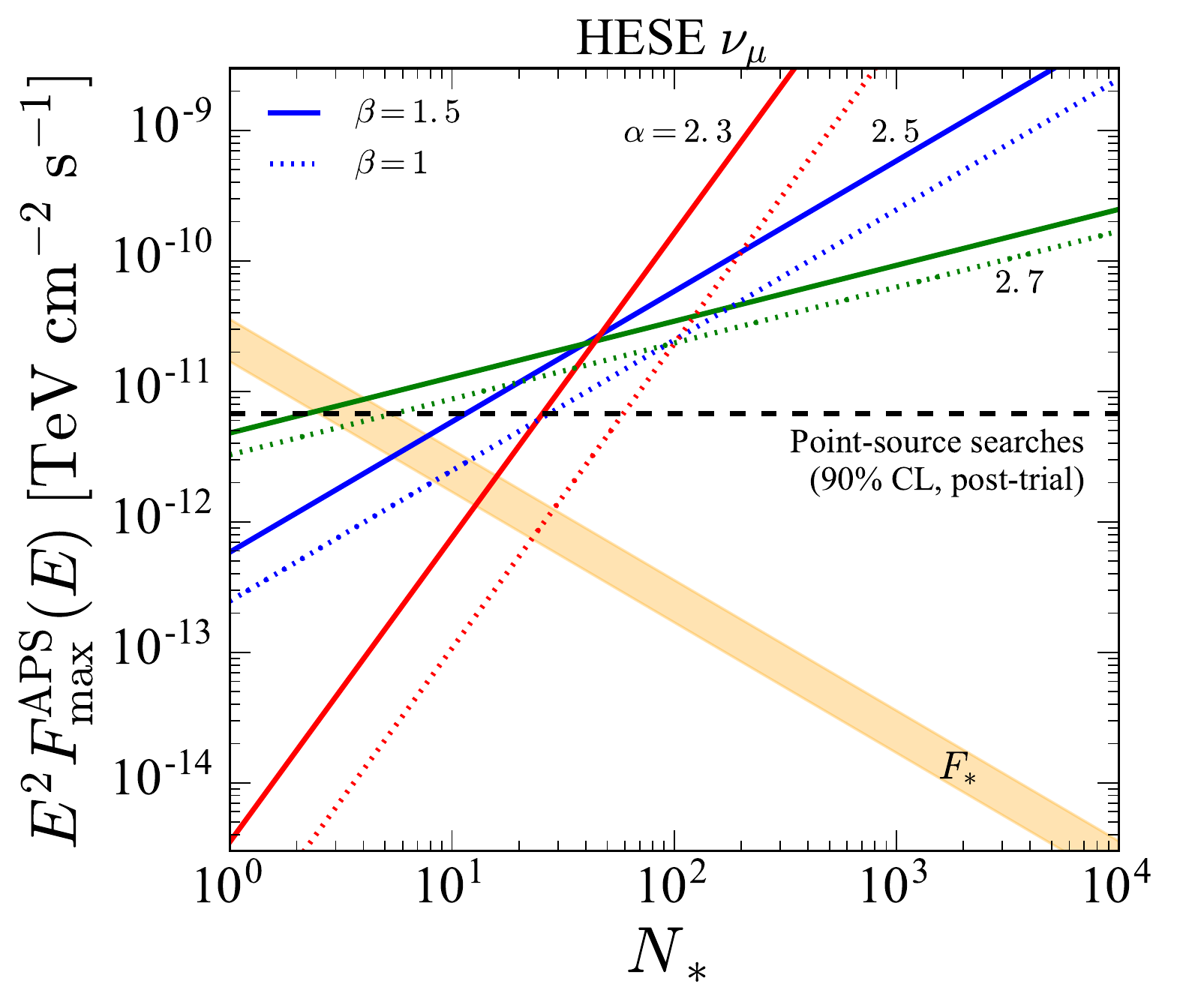}
  \caption{Upper limits (90\% CL) on the flux (per flavor) of the
  brightest source from the angular power spectrum, $F_{\rm max}^{\rm
  APS}$, as a function of the characteristic source number $N_*$
  by using the HESE dataset. The
  color code and line style are the same as in
  Fig.~\ref{fig:Fmax_CV}. Only the regions where $F_{\rm max}^{\rm APS}
  > F_*$ are valid as upper limits.}
  \label{fig:Fmax_APS}
 \end{center}
\end{figure}

Figure~\ref{fig:Fmax_APS} shows the $F_{\rm max}^{\rm APS}$ derived from HESE tracks, as a function of
$N_\ast$ and for different values of $\alpha$ and $\beta$. Values of $F_{\rm max}^{\rm APS}$ larger than the solid or dotted lines
are excluded, as the term due to the flux variance in Eq.~(\ref{eq:count
variance}) would have been detected in Ref.~\cite{IceCubeAnis}. For small values of $N_\ast$ (at most below $\sim$50, in the case with
$\alpha=2.3$ and $\beta=1$), the upper limits obtained here are more
stringent than those by the search for point-like
sources~\cite{IceCubePS}, let alone the one-source constraints considered earlier.
Note, however, that this upper limit is based on the assumption that
$F_{\rm max}^{\rm APS} > F_\ast$; otherwise the source flux distribution
would be proportional to $F^{-\beta}$ with a truncation at $F_\ast$ (see
Appendix~\ref{app:Case of a flat distribution}).

\subsection{Upgoing muon neutrinos}
\label{sec:numu}
It is possible to repeat the analysis above for high-energy upgoing
tracks, for which rather than requiring the interaction vertex be
contained one uses the Earth itself as a veto against atmospheric muon
backgrounds~\cite{Aartsen:2016xlq}. Above 300 TeV, it is possible to
estimate $C_\nu^N$ using the best-fit powerlaw models of astrophysical
flux $E^2 I_\nu=0.7\times
10^{-18}~\mathrm{GeV}~\mathrm{cm}^{-2}~\mathrm{sr}^{-1}~\mathrm{s}^{-1}$~\cite{Aartsen:2016xlq}
and the conventional atmospheric background $I_\nu \propto
E^{-3.7}$~\cite{Honda:2015fha}.
We adopt a sky coverage of $f_{\rm sky}\sim 0.5$, as well as the
energy-dependent effective area and construction-dependent livetimes of
the telescope from Ref.~\cite{Aartsen:2016xlq}. This corresponds to
$N_{\rm astro}\sim 56$ and $N_\mathrm{atm}\sim13$, and is consistent
with Fig.~1 from Ref.~\cite{Aartsen:2016xlq} where a cursory inspection
yields roughly 60 and 10 events above 300 TeV respectively. We adopt an
angular resolution of $\theta_{\rm psf} \sim 0.5^\circ$, better than for
the contained events of the previous section since the outermost optical
modules of IceCube are used to improve pointing rather than as a
veto. With these parameters, we obtain an upper limit on the Poisson
angular power spectrum of
\begin{equation}
E^4 C_{\nu,{\rm lim}}^P(E) = 2.1 \times 10^{-25} ~
\mathrm{TeV^2~cm^{-4}~s^{-2}~sr^{-1}},
\end{equation}
from uncontained, upgoing tracks above 300 TeV.

 \begin{figure}
 \begin{center}
  \includegraphics[width=8.5cm]{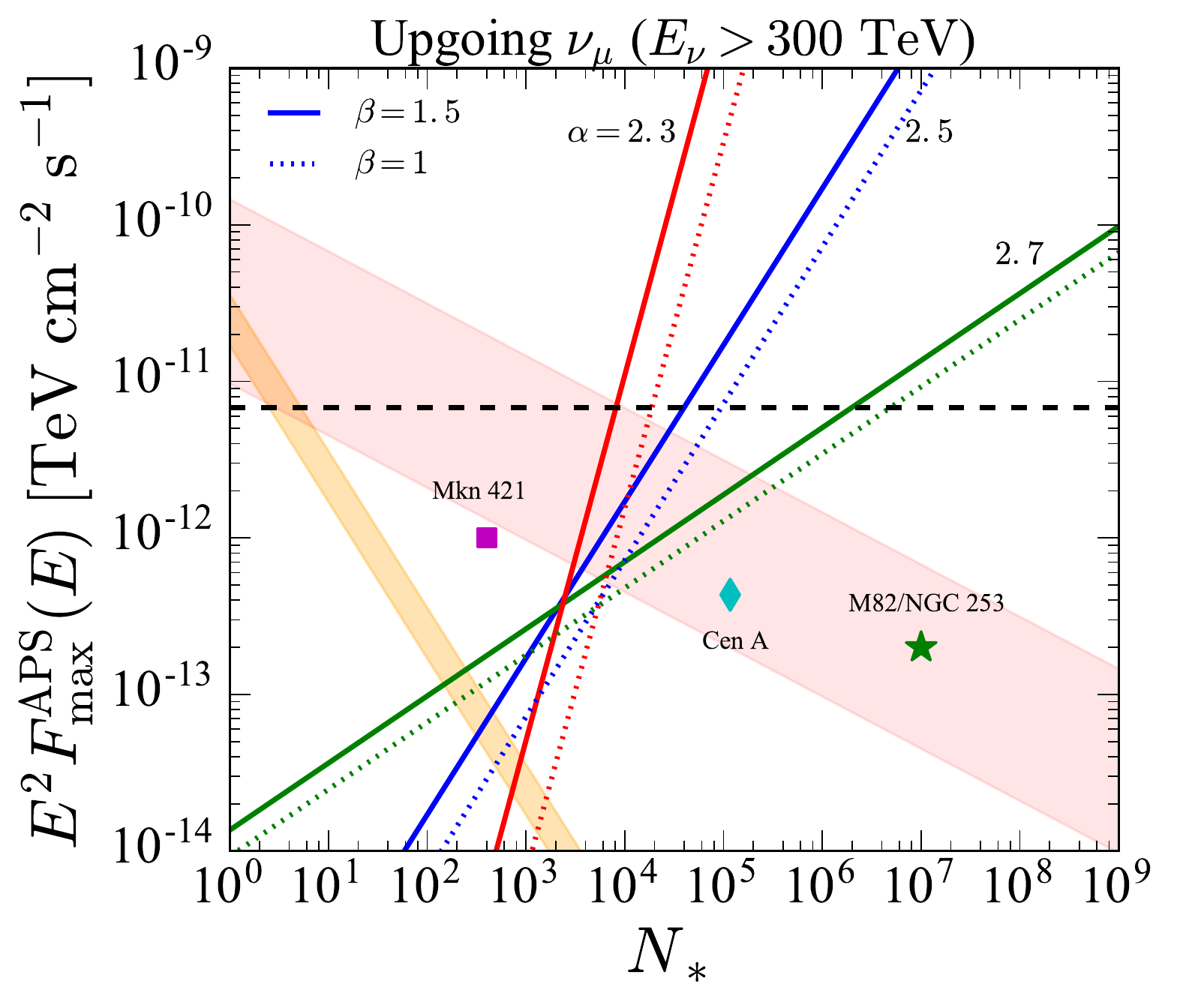}
  \caption{Upper limits (90\% CL) on the flux (per flavor) of the
  brightest source from the angular power spectrum, $F_{\rm max}^{\rm
  APS}$, as a function of the characteristic source number $N_*$
  by using the upgoing $\nu_\mu$ events above 300~TeV
  and assuming the current IceCube exposure~\cite{Aartsen:2016xlq}. The
  color code and line style are the same as in
  Fig.~\ref{fig:Fmax_CV}. Only the regions where $F_{\rm max}^{\rm APS}
  > F_*$ are valid as upper limits. The pink bank represents the region where 
  the brightest source is located at 90\% CL for given $N_\ast$, in the 
  Euclidean case with $(\alpha,\beta) = (2.5, 1.5)$. The purple square, blue
  diamond and green star are located at the expected neutrino flux for Mkn~412, Cen~A and M82 or NGC~253, for values of $N_\ast$ typical of blazars,
  radio galaxies and starburst galaxies, respectively (see text for details).}
  \label{fig:Fmax_APS_upgoing}
 \end{center}
\end{figure}

Figure~\ref{fig:Fmax_APS_upgoing} shows the $F_{\rm max}^{\rm APS}$
derived from upgoing tracks. These limits are many orders of
magnitude stronger than the limits from HESE as a result of the improved
angular resolution and the much larger exposure.
The ``pivot point'' for which the limit is independent of $\alpha$ is
also below the point-source searches.
In addition to these upper limits, we show the region containing the
brightest sources at 90\% CL derived in Sec.~\ref{sec:One-source
constraint}.
The absence of anisotropies will clearly constrain rare sources
better than point-source searches for $N_\ast \alt 10^4$.
Complementarily, for more abundant sources, the point-sources searches
do not cut into the brightest-source containment band, so we should not
expect (with 90\% CL) to have seen them yet anyway.
This is especially true if we expect multiple source populations to
contribute to this flux, since for populations contributing fractions
$k<1$ of the isotropic flux this band is even lower.
Even allowing for uncertainties in $(\alpha, \beta)$, these two
complemetary constraints (which rely only on the physically-motivated
assumption that source fluxes are power-law distributed) jointly place a
stronger constraint on the brightness of the brightest high-energy
neutrino source than current point-source searches.

\section{Application to known source populations}
\label{sec:Application to known source populations}

Although we aim to make our discussion as generic as possible, such that
it can be applied even to unknown classes of astrophysical sources that may contribute at high energies \cite{Ando:2017alx}, it is certainly of interest to discuss known source
populations in this context. We discuss mainly two source classes commonly thought to be the origin of the observed isotropic flux: BL Lacs \cite{Giacinti:2015pya,Righi:2016kio,Padovani:2015mba,Padovani:2016wwn} and starburst galaxies \cite{Murase:2013rfa,TAM,Anchordoqui:2014yva,Chang:2014hua,Senno:2015tra,ATZ}.

\subsection{Phenomenological representation}

The phenomenological parameterisation of a source population we introduced in
Sec.~\ref{sec:Formalism} can be summarised by the tuple $(\alpha,\beta,N_\ast)$.
The parameters for sources from the second catalog of hard Fermi sources (2FHL;
mostly BL Lacs) and starburst galaxies are $(\alpha,\beta,N_\ast)
\approx$ (2.5, 1.7, $6 \times 10^2$)~\cite{2FHL} and (2.5, 1.0,
$10^7$)~\cite{Feyereisen2016}, respectively.
These are estimated from their gamma-ray observations (with help of
infrared observations in the case of the starbursts) and assuming a
linear correlation between the gamma-ray and neutrino luminosities,
$L_\nu \propto L_\gamma$.
This is well supported for the case of starbursts, which emit neutrinos
through $pp$ interaction~\cite{TAM,ATZ}.
For the blazars emitting through $p\gamma$ interaction, on the other
hand, the relation between the gamma-ray and neutrino luminosities is
more complicated and model dependent, but see, e.g.,
Ref.~\cite{Padovani:2015mba} for a model of linear scaling.
Other cases with stronger dependence can also be accommodated with
similar parameters: e.g., $(\alpha, \beta, N_\ast) \approx (2.5, 1.25, 4
\times 10^2)$ for the BL Lacs with $L_\nu \propto L_\gamma^2$
scaling~\cite{Tavecchio2015}, and $(2.3, 0.9, 1.5\times 10^2)$ for the
flat-spectrum radio quasars with $L_\nu \propto
L_\gamma^{1.5}$~\cite{MID}.
See Appendix~\ref{app:Examples of extreme blazar models} for more
discussions for these cases.

With these parameters, Figs.~\ref{fig:Fmax_CV} and \ref{fig:Fmax_APS_upgoing}
show that the 90\% CL upper limits on the flux $F_{\rm max}$ of the brightest high-energy neutrino source, are
\begin{equation}
F_\mathrm{max}^\mathrm{BL~Lac} \sim 10^{-13}~\mathrm{TeV~cm^{-2}~s^{-1}},
\label{eq:BLLAC_UL}
\end{equation} for the 2FHL sources, based on the angular power spectrum
constraint, and 
\begin{equation}
F_\mathrm{max}^\mathrm{starburst} \sim 6 \times 10^{-13}~\mathrm{TeV~cm^{-2}~s^{-1}},
\end{equation} for the starbursts, based on the one-source
constraint. Recall that these upper limits are on the flux per flavor of a population contributing a fraction $k=1$ of the observed astrophysical flux, assuming an $E^{-2}$ spectrum, and requiring (for the former constraint) an absence of detectable anisotropies.

\subsection{Physical representation}

Up to this point, we considered $\alpha$, $\beta$ and $N_\ast$ as free
parameters.
Another complementary representation is to use more physical quantities
such as luminosity $L_\nu$ and density $n_s$ of the sources, although
the discussion will be model dependent.
The latter approach was taken in, e.g., Refs.~\cite{Silvestri:2009xb,
Ahlers:2014ioa, MuraseWaxman}, where sources were assumed to have the
same luminosity.
These two representations can be converted from one to the other through
\begin{eqnarray}
 F_\ast &\simeq& 10^{-18}
  \left(\frac{L_\nu}{10^{40}~\mathrm{erg~s^{-1}}}\right)
   ~ \mathrm{TeV~cm^{-2}~s^{-1}},
  \label{eq:Fstar} \\
 N_\ast &\simeq& 3 \times 10^7 k
  \left(\frac{\eta_1}{4}\right)^{-1}
  \left(\frac{L_\nu}{10^{40}~\mathrm{erg ~ s^{-1}}}\right)^{-1},
  \label{eq:Nstar} \\
 F_{\rm max}^{\rm 1s} &\simeq& 3 \times 10^{-13}
  \left(\frac{n_s}{10^{-5}~\mathrm{Mpc^{-3}}}\right)^{2/3}
  \nonumber\\&&{}\times
  \left(\frac{L_\nu}{10^{40}~\mathrm{erg~s^{-1}}}\right)
  ~\mathrm{TeV~cm^{-2}~s^{-1}},
\end{eqnarray}
in the case of $\alpha=2.5$.
Typically $(n_s, L_\nu) = (10^{-5}~\mathrm{Mpc^{-3}}, 2\times
10^{40}~\mathrm{erg~s^{-1}})$ and $(10^{-7}~\mathrm{Mpc^{-3}},
2\times 10^{44}~\mathrm{erg~s^{-1}})$ for the starbursts and
BL Lacs, respectively~\cite{MuraseWaxman}.
However, these relations apply only to mono-luminous case as was
studied in the literature.
See Appendix~\ref{sec:Relation to source density and luminosity} for
their derivation and more discussions.

In Fig.~\ref{fig:Fmax_APS_upgoing} (and those that follow), we show reference fluxes of some
well known sources for each class: Mkn~421 for the BL Lac blazars and
M82 or NGC~253 for the starbursts.
Mkn~421 is predicted to have a flux around $10^{-12} ~
\mathrm{TeV~cm^{-2}~s^{-1}}$ in a model of Ref.~\cite{Tavecchio2015}.
For M82 and NGC~253, we estimate the neutrino luminosity from the
gamma-ray luminosity of these sources~\cite{M82NGC253}, and then by
converting to the neutrino luminosity assuming $pp$
interaction~\cite{TAM}.
In addition, we show predicted neutrino flux from the most promising
radio galaxy, Cen~A, assuming production from $pp$
interaction~\cite{Hooper2016b}.
We assume that these sources are drawn from a population of emitters with
the same luminosity.
Thus, the number of sources can be estimated by Eq.~(\ref{eq:Nstar})
with $k = 1$, $\eta_1 = 4$, and typical neutrino luminosity for this
population found in Ref.~\cite{MuraseWaxman}.

\subsection{Discussion}

All these sources fall within the 90\% region of $F_{\rm max}$ predicted
with the one-source argument with $(\alpha,\beta)=(2.5,1.5)$ (shown as a
red band in Fig.~\ref{fig:Fmax_APS_upgoing}) and so a source from any of these populations is plausibly the brightest neutrino source.
A slight tension exists for Mkn~421, but
Ref.~\cite{Tavecchio2015} predicts several more BL Lacs with similar
flux such as PKS~2155-304, and the tension might go away when using a
fraction $k<1$ for the blazars.
The 90\% containment band for $N_\ast \approx 10^7$ is an order of
magntidue below the point-source constraint, suggesting it would be
unlikely to identify starburst galaxies amongst the brightest neutrino
sources.
This result is consistent with the analyses in
Refs.~\cite{Feyereisen2016, MuraseWaxman}.

The angular power spectrum is especially constraining for rare sources such as blazars.
The upper limit, Eq.~(\ref{eq:BLLAC_UL}), is nearly an order of
magnitude lower than the 90\% containment band for $N_\ast \approx
6\times 10^2$ and the predicted neutrino flux of Mkn~421.
The isotropy of the upgoing $\nu_\mu$ flux, if confirmed with the
current IceCube exposure, will force us to abandon the assumption that
they contribute a fraction $k=1$ of the high-energy neutrino flux.
This not only eases the aforementioned one-source tension for Mkn~421,
but furthermore is consistent with the analysis in
Ref.~\cite{Glusenkamp:2015jca}.

\section{Prospects for the future}
\label{sec:Prospects for the future}

In this section, we forecast the prospects for studying the flux of the brightest source with the next generation of neutrino telescopes, under the assumption that anisotropy searches will continue to yield null results in the future.

The angular power spectrum will become much more powerful for IceCube-Gen2~\cite{IceCubeGen2} and KM3NeT~\cite{KM3NeT}.
This is because of the strong dependence of $F_{\rm max}^{\rm APS}$ on
$C_{\nu,{\rm lim}}^P$ from Eq.~(\ref{eq:Fmax_APS}), where $C_{\nu,{\rm
lim}}^P$ improves with exposure as described by Eq.~(\ref{eq:deltaCl}).
For Euclidean sources ($\alpha=2.5$), the upper limit improves
quadratically with exposure: $F_{\rm max}^{\rm APS} \propto \mathcal
E^{-2}$.
The anticipated tenfold increase in exposure expected for
IceCube-Gen2 with respect to the current IceCube~\cite{IceCubeGen2} will yield hundredfold improvement on $F_{\rm max}^{\rm APS}$ if the observed angular
power spectrum remains consistent with isotropy, before even accounting
for any improvements in angular resolution.

 \begin{figure}
 \begin{center}
  \includegraphics[width=8.5cm]{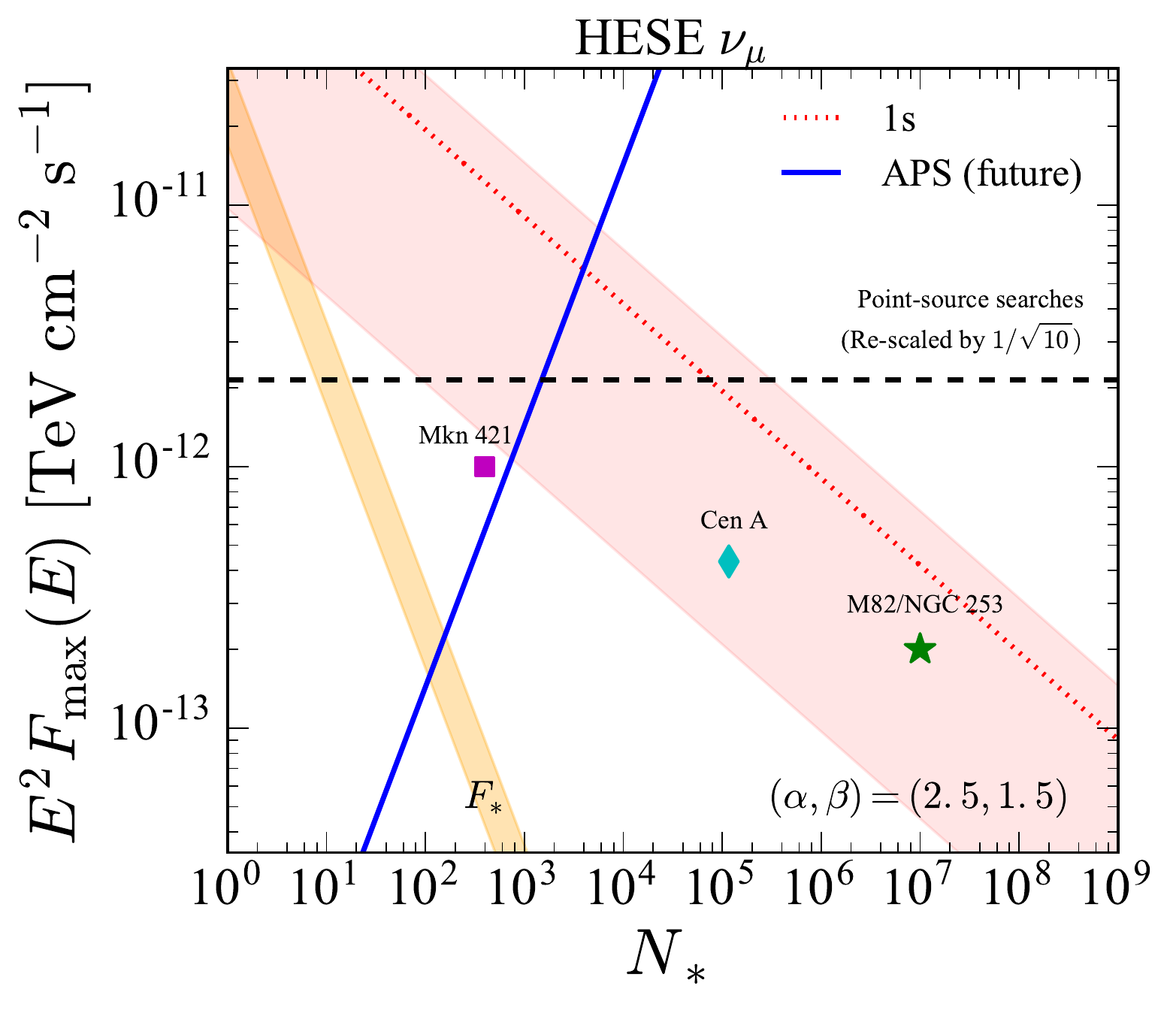}
  \caption{Projected 90\% CL upper limits from angular power spectrum
  (solid) and one-source limits (dotted) as a function of $N_\ast$,
  for contained track events, assuming
  $(\alpha,\beta)=(2.5,1.5)$. Projections for both KM3NeT and
  IceCube-Gen2, being coincidently the same, are shown as a solid
  line. The dashed
  horizontal line represents the upper limit from the search for
  point-like sources~\cite{IceCubePS} after scaled down by a factor of
  $1/\sqrt{10}$. The dotted line represents the 90\% CL one-source upper
  limits, and the red region shows where the flux of the brightest
  source is located at 90\% CL in the case of Euclidean sources.}
 \label{fig:Fmax_future_HESE}
 \end{center}
 \end{figure}
 
  \begin{figure}
 \begin{center}
  \includegraphics[width=8.5cm]{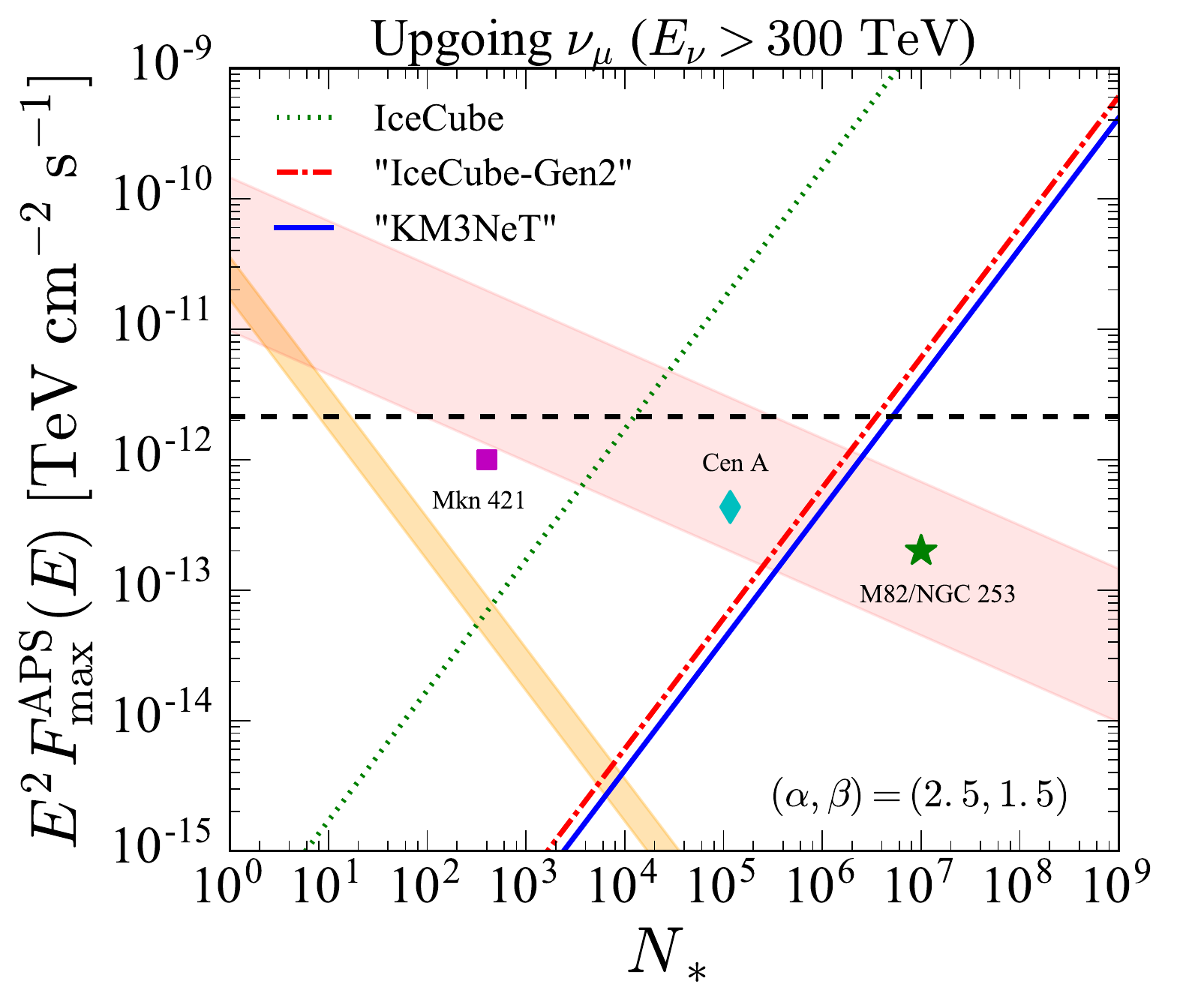}
  \caption{Projected 90\% CL upper limits from angular power spectrum
  as a function of $N_\ast$, if the high-energy neutrino sky remains
  isotropic after using detectors
  similar to KM3NeT (solid) and IceCube-Gen2 (dot-dashed), assuming
  $(\alpha,\beta)=(2.5,1.5)$. The dotted is for the current IceCube
  configuration as in Fig.~\ref{fig:Fmax_APS_upgoing}. See
  Table~\ref{tab:forecastparams} for detector configurations. The dashed
  horizontal line represents the upper limit from the search for
  point-like sources~\cite{IceCubePS} after scaled down by a factor of
  $1/\sqrt{10}$, and the red region shows where the flux of the brightest
  source is located at 90\% CL in the case of Euclidean sources.}
 \label{fig:Fmax_future_upgoing}
 \end{center}
 \end{figure}

Figures~\ref{fig:Fmax_future_HESE} and \ref{fig:Fmax_future_upgoing}
summarize future prospects for upper limits on the flux of the brightest
source, drawn from a population described by $\alpha=2.5$ and
$\beta=1.5$, with an improved track angular resolution and larger
exposures than acheived today (cf. Table~\ref{tab:forecastparams}).
For comparison, we scale down the upper limit from the search of
point-like sources by a factor of $1/\sqrt{10}$, assuming that these
analyses are already background limited; the value of $F_{\rm max}^{\rm
1s}$ from the one-source constraints remains unchanged.

\begin{table}
\caption{Parameters used in forecasts of $F_\mathrm{max}^\mathrm{APS}$
 in the scenario the astrophsyical flux remains consistent with
 isotropy. Exposures are shown normalized to the current IceCube
 searches in Refs.~\cite{IceCubeScience} (HESE) and
 \cite{Aartsen:2016xlq} (upgoing $\nu_\mu$). The equivalent livetimes
 and the angular resolutions are estimated from
 Refs.~\cite{IceCubeGen2,KM3NeT}.}
\medskip
\centering
\begin{tabular}{llccc}
\hline\hline
Detector & Strategy & $\mathcal{E}/\mathcal{E}_{\rm today}$ & livetime & $\theta_\mathrm{psf}$ (tracks) \\
\hline
IceCube& HESE & 1 & 4 yr & $1^\circ$\\
&upgoing $\nu_\mu$ & 1 & 6 yr & $0.5^\circ$\\
IceCube-Gen2 & HESE & 10 & 8 yr & $0.5^\circ$ \\
&upgoing $\nu_\mu$ &10& 12 yr & $0.3^\circ$ \\
KM3NeT & HESE & 4 & 8 yr & $0.2^\circ$ \\
& upgoing $\nu_\mu$ & 4 & 12 yr & $0.1^\circ$\\
\hline\hline
\end{tabular}
\label{tab:forecastparams}
\end{table}

In future HESE-like analyses, the limits on $F_{\rm max}$ from the
angular power spectrum from IceCube-Gen2 and KM3NeT (summarized in
Fig.~\ref{fig:Fmax_future_HESE}) will outperform point-source searches
only if the isotropic flux is due to individually bright sources rarer
than $N_\ast \alt 10^3$.
In this hypothetical nondetection scenario, the parameter space
associated to blazars would not be constrained much better than it is
today using upgoing events (cf. Fig.~\ref{fig:Fmax_APS_upgoing}),
due to limited improvements in exposure, as well as in
angular resolution.

Constraint prospects for future analyses of upgoing (uncontained) tracks
are summarized in Fig.~\ref{fig:Fmax_future_upgoing}.
In the pessimistic case studied here of a continued nondetection of
anisotropy or point sources, KM3NeT and IceCube-Gen2 would
(independently and with high significance) rule out a blazar
contribution to the high-energy neutrino flux observed today.
The angular power spectrum from the next generation of neutrino
telescope also has the potential to constrain radio galaxies.
Indeed, the upper limits for $N_\ast \approx 10^5$ would reach down to
$5\times 10^{-14}~ \mathrm{TeV~cm^{-2}~s^{-1}}$ by the time these
experiments are decommisioned, well below their neutrino flux
anticipated from $pp$ interactions \cite{Hooper2016b}.
In both HESE and upgoing track analyses, the one-source constraint will
still be the most stringent on the population of starburst galaxies,
suggesting that it will still be unlikely for the neutrino telescopes to
detect them (see also Refs.~\cite{Feyereisen2016, MuraseWaxman}).

These forecast clearly shows that in the future, if the high-energy
neutrino sky remains consistent with isotropy, the angular power
spectrum will provide much stronger upper limits on the flux of the
brightest neutrino source than point-source searches.
It also suggests (by comparison with Fig.~\ref{fig:Fmax_APS_upgoing})
that if sources are not discovered individually in the near future, they
will likely be discovered statistically through the angular power
spectrum first.
Indeed, due to the respective $\sqrt{\mathcal{E}}$ and $\mathcal{E}^2$
scalings of the point-source search and the APS, a statistical discovery
becomes increasingly likely the longer point sources are not
discovered.

\section{Conclusions}
\label{sec:Conclusions}

To conclude, we discussed two constraints on the flux of the brightest
neutrino source in the sky, $F_{\rm max}$, and how they relate to (or
improve on) the null results of the current anisotropy and point-source
searches.
The one-source limit on $F_{\rm max}$ manages to reach quite low values,
more than one order of magnitude below the existing upper limits based
on the search for individual point-like sources in the case of abundant
source population such as starburst galaxies.
The other approach is based on constraining the variance of source flux 
distribution (or equivalently, the Poisson angular power spectrum).
These upper limits are more powerful for rare source classes, 
providing complementary information in the case that no source is
detected.
In particular, analysis of upgoing $\nu_\mu$ track
events with the current IceCube exposure already has a potential to rule
out the scenario of blazar-domination for the diffuse neutrino flux.
In addition, the limits based on the angular power spectrum will become
more powerful for the next generation of neutrino telescopes.
The combination of the two strategies proposed here provide a very
efficient way of answering the question: ``How bright can the brightest
neutrino source be?''

\acknowledgments
We thank Markus Ahlers, John Beacom, Kohta Murase, and an anonymous
referee for helpful comments and discussions on the manuscript.
This work was supported by the Netherlands Organization for Scientific 
Research (NWO) through a Vidi Grant.

\appendix

\section{Flux upper limits of the brightest source from point-source
 searches}
\label{app:PS}

The point-source flux upper limits are dependent on declination
$\delta$~\cite{IceCubePS}.
In this paper, however, we are interested in a single value of the flux
of the {\it brightest} neutrino source.
Here we shall discuss how we estimate this flux.

Suppose $F_{\rm max}$ is the flux of the single brightest source
somewhere in the sky.
Above the flux corresponding to the point-source upper limit
$F_{\rm lim}(\delta)$ at the declination $\delta$ (where $F_{\rm
lim}(\delta) < F_{\rm max}$), there will be on average $[F_{\rm
lim}(\delta) / F_{\rm max}]^{-\alpha + 1}$ sources from the full sky.
The number of sources above this threshold in a declination bin
$\Delta\delta$ is therefore $\Delta N_s = [F_{\rm lim}(\delta) / F_{\rm
max}]^{-\alpha + 1} \Delta\sin\delta / 2$.
We then assign a probability $p$ of finding no source brighter than the
current point-source upper limits anywhere in the sky, through the
Poisson statistics, as
\begin{equation}
 p = \exp
  \left[-\frac{1}{2}\int_{F_{\rm lim}(\delta) < F_{\rm max}}
   d\sin \delta \left(\frac{F_{\rm lim}(\delta)}{F_{\rm
		 max}}\right)^{-\alpha+1}\right].
\end{equation}
By using post-trial 90\% CL upper limits $F_{\rm lim}(\delta)$ from
Ref.~\cite{IceCubePS}, $\alpha = 2.5$, and $p = 0.1$, we solve this
equation for $F_{\rm max}$, and obtain $E^2 F_{\rm max}(E) = 6.8\times
10^{-12}~ \mathrm{TeV ~ cm^{-2} ~ s^{-1}}$.

\section{Relation to source density and luminosity}
\label{sec:Relation to source density and luminosity}

We shall characterize a source population by its local number density
$n_s$ and the neutrino luminosity $L$.
Assuming that they are distributed in a local volume where cosmological
effects can be neglected, the number of sources that give fluxes greater
than $F$ is then $n_s$ multiplied by a volume with a radius $r = (L/4\pi
F)^{1/2}$:
\begin{equation}
 N_s(>F) = \frac{n_s L^{3/2}}{6\sqrt{\pi} F^{3/2}},
\end{equation}
from which one can derive $P_1(F) = d\ln N_s / dF \propto F^{-5/2}$.
Taking the luminosity distribution into account, we replace $L^{3/2}$
with its average over the luminosity function $\langle L^{3/2}
\rangle$.

Then, as above, the one-source limit is obtained with $p = 1-
\exp[-N_s(>F_{\rm max}^{\rm 1s})]$, which reads
\begin{eqnarray}
 F_{\rm max}^{\rm 1s} &=&
  \left(\frac{-n_s \langle L^{3/2} \rangle}{6\sqrt{\pi}\ln (1-p)}\right)^{2/3} 
  \nonumber\\&=&
  3\times 10^{-13}
  \left(\frac{n_s}{10^{-5}~\mathrm{Mpc^{-3}}}\right)^{2/3}
  \nonumber\\&&{}\times
  \left(\frac{\langle
   L^{3/2}\rangle^{2/3}}{10^{40}~\mathrm{erg~s^{-1}}}\right)
  ~\mathrm{TeV~cm^{-2}~s^{-1}}.
\end{eqnarray}
Here we again choose $p = 0.1$.

The break of the flux distribution at its characteristic flux $F_\ast$
happens when the cosmological expansion comes into play.
Although this is dependent on how the source density evolves as a
function of redshift $z$ and one needs to fully compute $P_1(F)$ in
order to be more precise (e.g., \cite{Feyereisen2016}), here we simply
approximate that the transition happens at $z = 1$: $F_\ast = \langle
L\rangle/[4\pi d_L^2(z=1)]$, where $d_L$ is the luminosity distance.
We then obtain $N_\ast$ using Eq.~(\ref{eq:nuintensity}) by replacing
measured $I_\nu$ with $k I_\nu$, where $k (<1)$ is a fractional
contribution to the measured intensity from the source population.
They are
\begin{eqnarray}
 F_\ast &\simeq& 10^{-18}
  \left(\frac{\langle L\rangle}{10^{40}~\mathrm{erg~s^{-1}}}\right)
  ~\mathrm{TeV~cm^{-2}~s^{-1}},
  \\
 N_\ast &\simeq& 7\times 10^7 k
  \left(\frac{\eta_1}{4}\right)^{-1}
  \left(\frac{\langle
   L\rangle}{10^{40}~\mathrm{erg~s^{-1}}}\right)^{-1}.
\end{eqnarray}

If the sources are mono-luminous (i.e., the luminosity function is
sharply peaked at some value) as is often assumed in the
literature~\cite{Silvestri:2009xb, Ahlers:2014ioa, MuraseWaxman}, then
all these quantities are determined once $n_s$ and $L$ are both given.
In this case, by equating $F_{\rm max}^{\rm 1s}$ with the upper limits
from the point-source searches, one can place an exclusion line on the
$(n_s, L)$ plane.
In general, however, the luminosity function can range widely, and if
it is flatter than $L^{-2.5}$, then $\langle L^{3/2} \rangle$ and hence
$F_{\rm max}^{\rm 1s}$ are mainly sensitive to the upper cutoff of the
luminosity function.
Such a behavior in the tail region of the luminosity function is
typically found for the blazars in the gamma rays~\cite{BLLacLF,
FSRQLF}, and expected in neutrinos too (see the next section).

\section{Examples of blazar models with flat luminosity distribution}
\label{app:Examples of extreme blazar models}

If there is a linear correlation between the neutrino and gamma-ray
luminosities, $L_\nu \propto L_\gamma$, then one can adopt
well-established flux distribution from the gamma-ray measurements such
as Ref.~\cite{2FHL}.
However, if the neutrinos are produced by the $p\gamma$ interaction, and
if its opacity is dependent of the gamma-ray luminosity, then the
scaling can be different from linear.
Here, we take recent examples that predict a stronger correlation,
$L_\nu \propto L_\gamma^r$, where $r > 1$.
There are models of BL Lacs with $r = 2$~\cite{Tavecchio2015} and
flat-spectrum radio quasars (FSRQs) with $r = 1.5$~\cite{MID}.
These kinds of dependence yield a flat distribution of the neutrino
luminosities.

The purpose of this section is to obtain the flux distribution starting
from the gamma-ray luminosity function, $dn_s / dL_\gamma$.
The neutrino intensity is
\begin{equation}
 E^2I_\nu(E) = \int dz \frac{d^2V}{dzd\Omega}\int dL_\gamma
  \frac{dn_s}{dL_\gamma} E^2 F_\nu (E, L_\gamma, z),
  \label{eqapp:intensity}
\end{equation}
where $V$ is the comoving volume, $E^2 F_\nu (E) \propto L_\gamma^r /
d_L^2$, and $d_L$ is the luminosity distance corresponding to the
redshift $z$.
We adopt the luminosity functions from Ref.~\cite{FSRQLF} for FSRQs and
Ref.~\cite{BLLacLF} for BL Lacs, but with the cutoff of $L_\gamma <
10^{46}~\mathrm{erg~s^{-1}}$ for the latter case~\cite{Tavecchio2015}.
Using the measured intensity $E^{2} I_\nu(E) = (0.84 \pm 0.3) \times
10^{-11}$~TeV~cm$^{-2}$~s$^{-1}$~sr$^{-1}$~\cite{IceCubeICRC}, we solve
Eq.~(\ref{eqapp:intensity}) to obtain the constant of proportionality of
the scaling relation $E^2 F_\nu(E) \propto L_\gamma^r / d_L^2$.
Then, the flux distribution is calculated as
\begin{equation}
 \frac{dN_s}{dF_\nu} = (4\pi)^2 \int dz \frac{d^2V}{dzd\Omega}d_L^2
 \frac{dn_s}{dL_\gamma}\frac{dL_\gamma}{dL_\nu},
\end{equation}
where both $L_\gamma$ and $L_\nu$ are now functions of $F_\nu$ and $z$.

\begin{figure}
 \begin{center}
  \includegraphics[width=8.5cm]{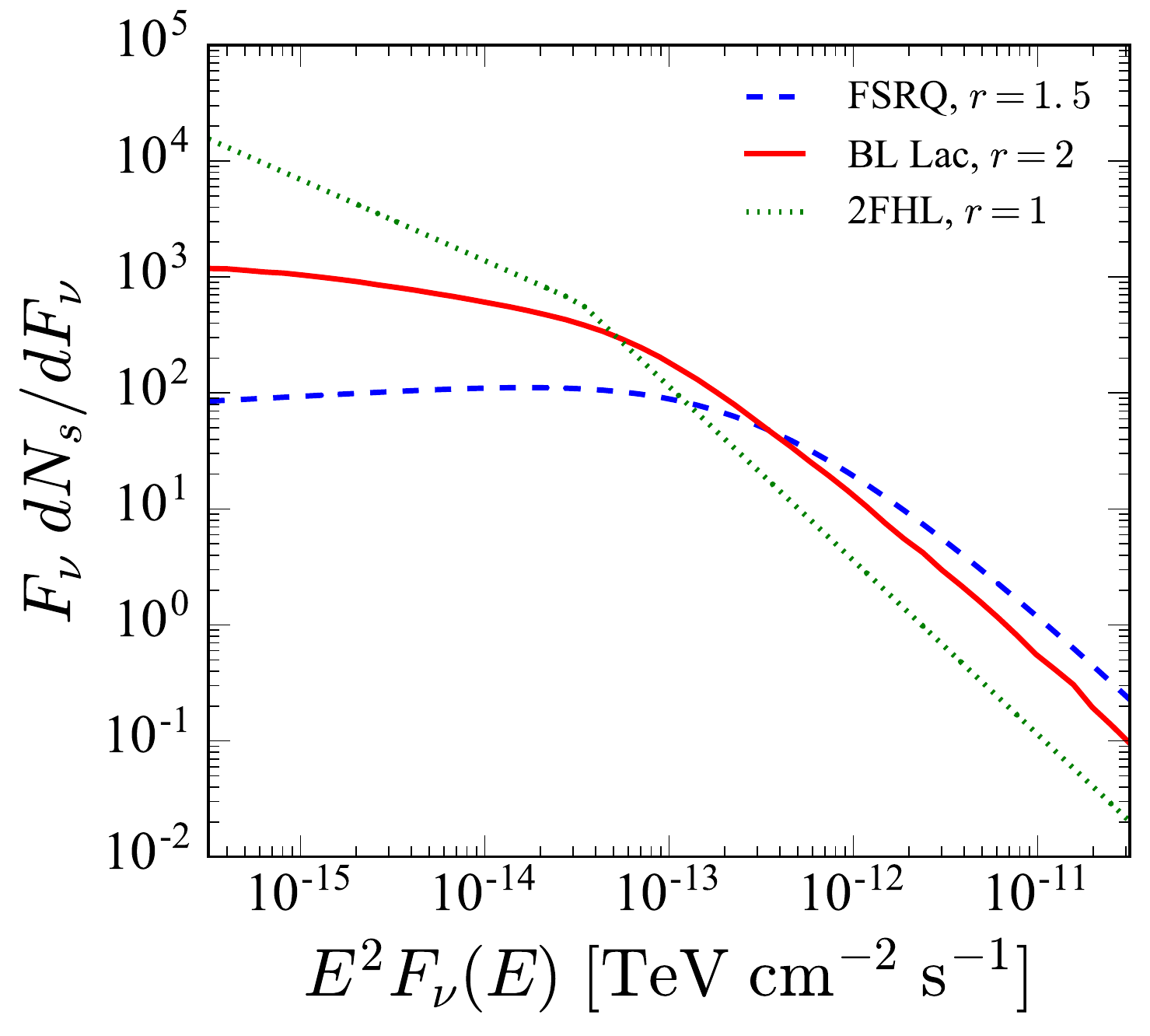}
  \caption{Flux distribution $F_\nu dN_s/dF_\nu$ for the BL Lac model
  with $L_\nu \propto L_\gamma^2$~\cite{Tavecchio2015} and the FSRQ
  model with $L_\nu \propto L_\gamma^{1.5}$~\cite{MID}, compared with
  the 2FHL distribution~\cite{2FHL} assuming linear scaling, $L_\nu
  \propto L_\gamma$.}
  \label{fig:FdNdF_blazar}
 \end{center}
\end{figure}

Figure~\ref{fig:FdNdF_blazar} shows $F_\nu dN_s/dF_\nu$ for both the
models of BL Lac~\cite{Tavecchio2015} and FSRQs~\cite{MID}, and compare
the one of 2FHL~\cite{2FHL} assuming a linear scaling $r = 1$.
All these models are normalized such that each of them can explain the
measured diffuse neutrino intensity entirely.
This shows that our phenomenological model based on a simple assumption
of the broken power law, with $2 < \alpha < 3$ at high-flux tail, indeed
captures the overall behavior of the flux distribution, predicted with a
realistic gamma-ray luminosity function and even in combination with
very strong scaling relations between the neutrino and gamma-ray
luminosities.

\section{Case of a flat distribution}
\label{app:Case of a flat distribution}

Here we address the case where $1 < \alpha < 2$ and $\alpha > \beta$.
As seen in the previous section, this case is very difficult to realize,
but in order to make our discussion fully generic, we study it.
One example of models that can potentially feature a flat tail in the
flux distribution is the case where one expects virtually no source in
the local volume with $z < 1$.
This is again extremely hypothetical and even unrealistic, because even
for starburst galaxies, while the redshift evolution is very steep (the
luminosity density evolves as $\propto (1+z)^3$ or steeper~\cite{TAM}),
the flux distribution has the Euclidean tail,
$F^{-2.5}$~\cite{Feyereisen2016}.

\begin{figure}
 \begin{center}
  \includegraphics[width=8.5cm]{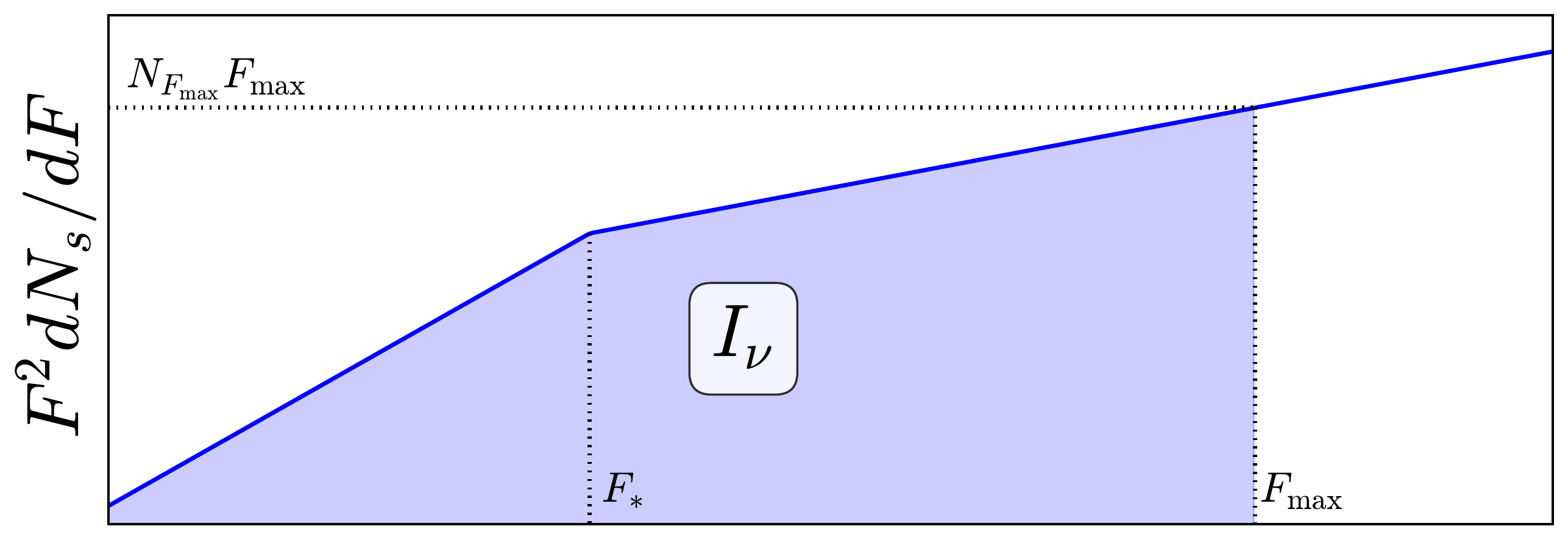}
  \includegraphics[width=8.5cm]{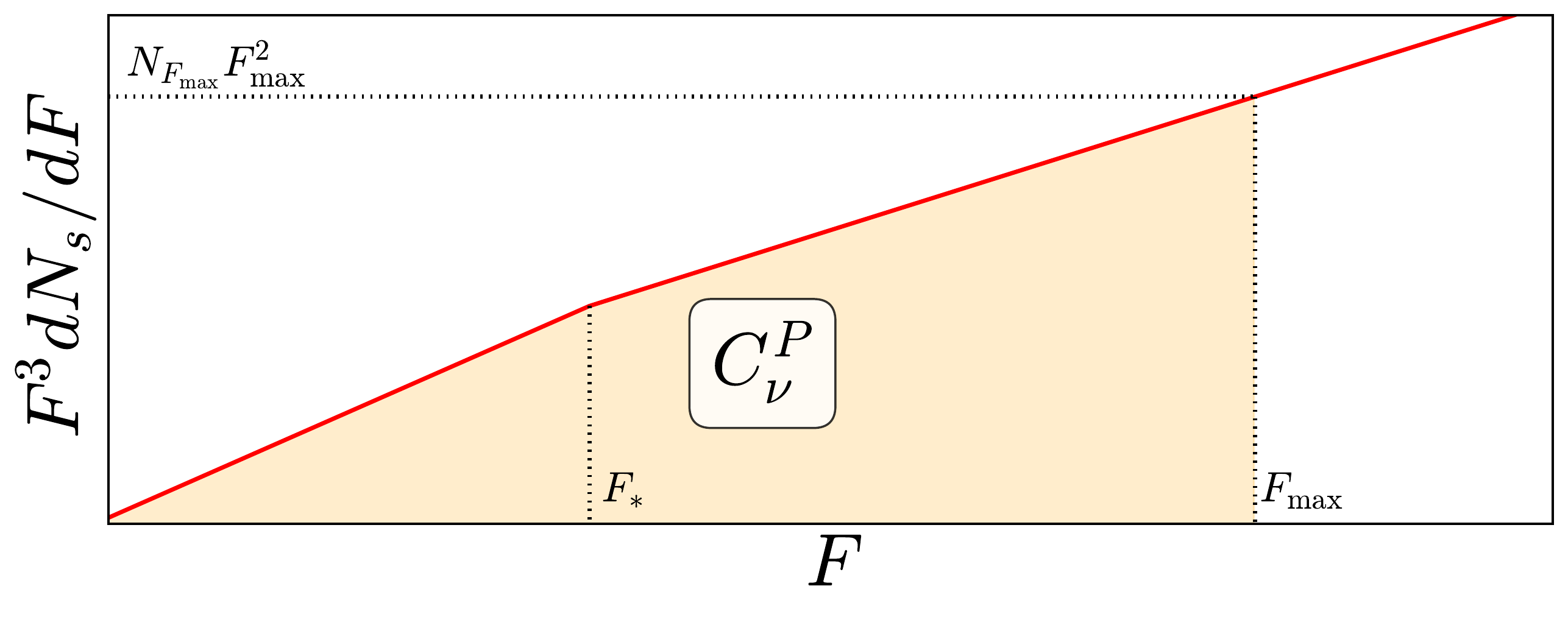}
  \caption{The same as the middle and bottom panels of
  Fig.~\ref{fig:dNdF}, but for $\alpha < 2$.}
  \label{fig:dNdF_flatdist}
 \end{center}
\end{figure}

\begin{figure}
 \begin{center}
  \includegraphics[width=8.5cm]{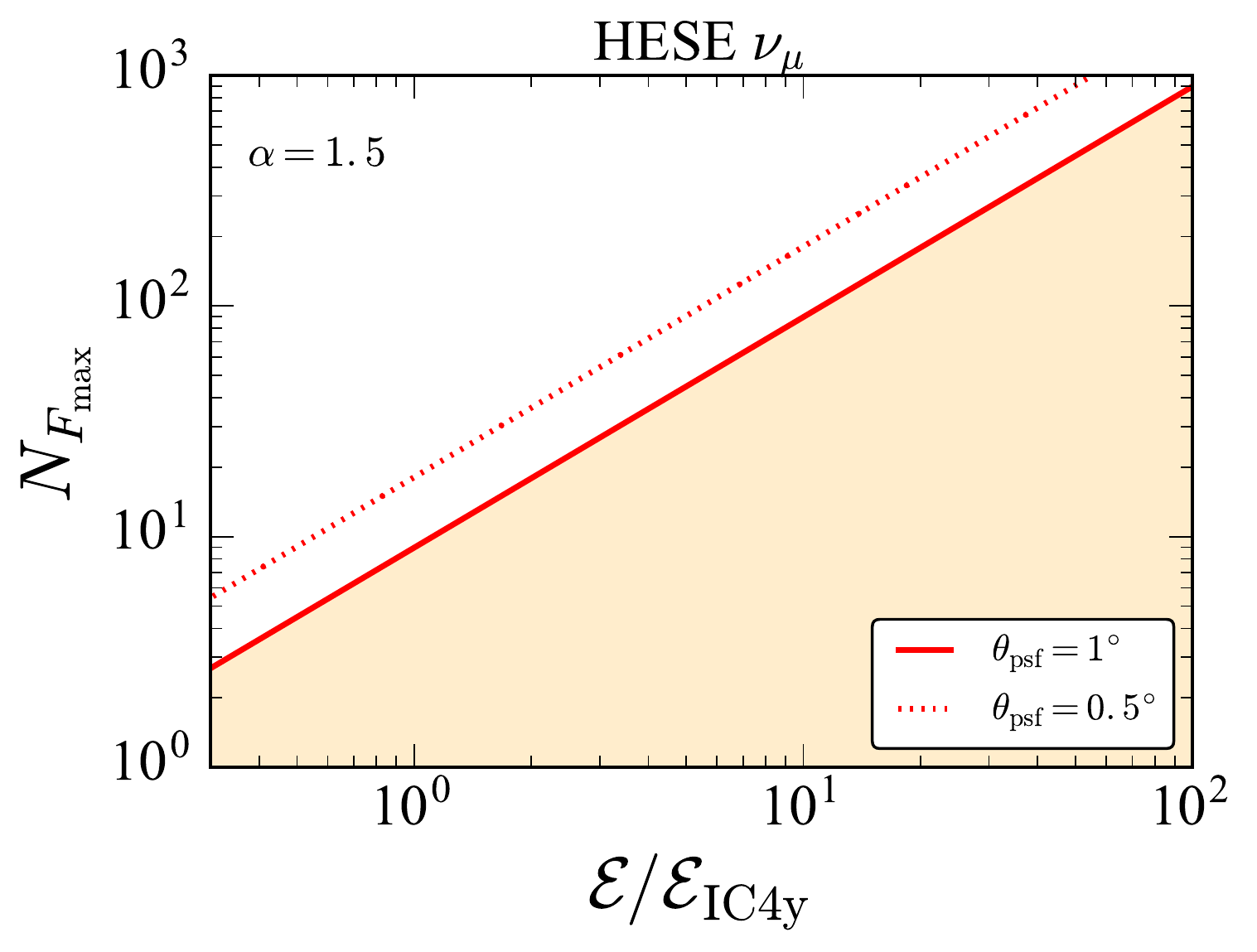}
  \includegraphics[width=8.5cm]{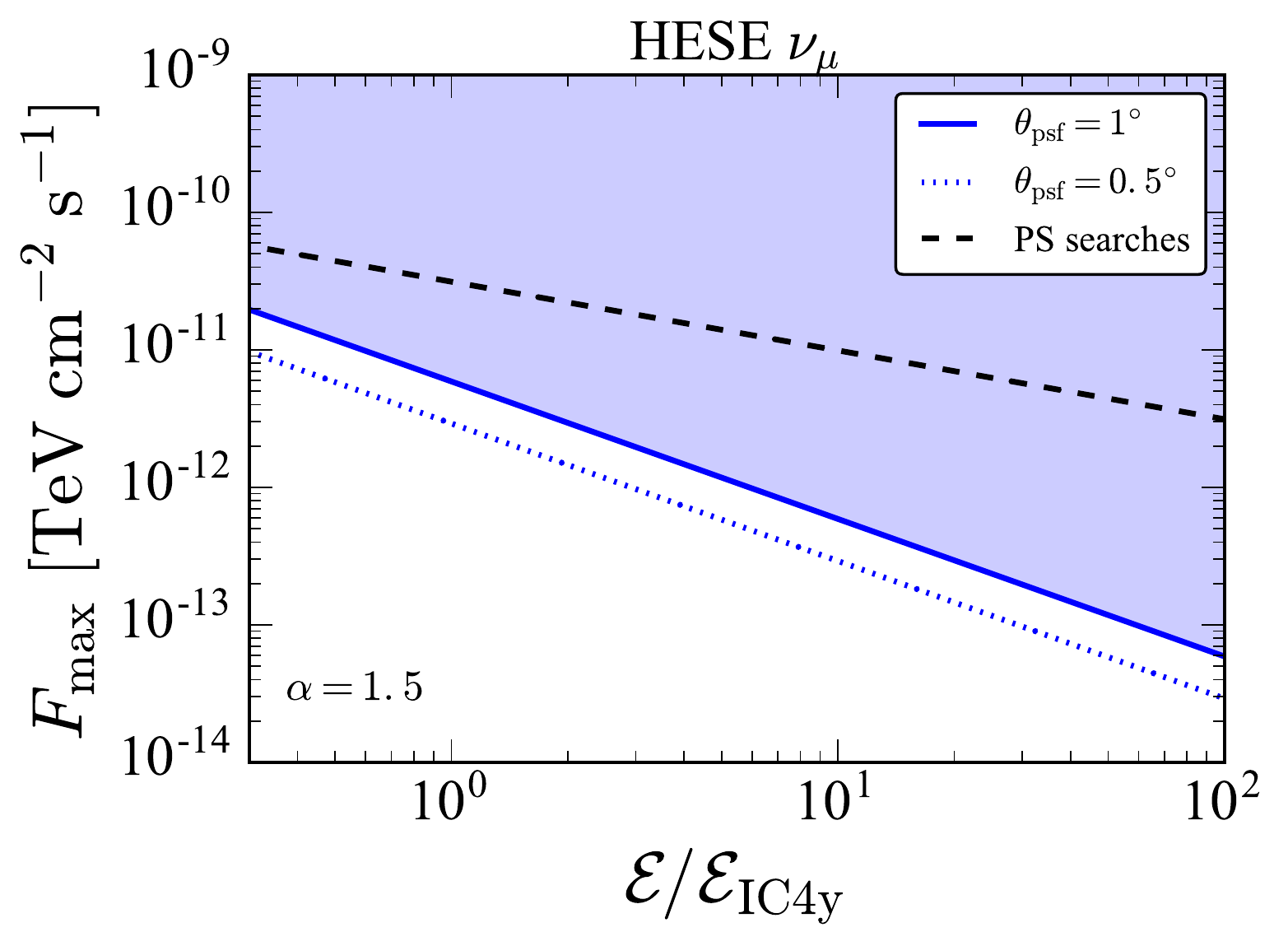}
  \caption{Lower limits on $N_{F_{\rm max}}$ (top) and upper limits on
  $F_{\rm max}$ (bottom) as a function of exposure normalized to that of
  4-year IceCube $\mathcal E_{\rm IC4y}$, from the angular power
  spectrum measurements, in the
  case of $\alpha = 1.5$. Solid and dotted lines correspond to angular
  resolutions of $\theta_{\rm psf} = 1^{\circ}$ and $0.5^{\circ}$,
  respectively. The dashed line in the bottom panel is the upper limits
  from the point-source searches (Appendix~\ref{app:PS}), extrapolated
  as $\mathcal E^{-1/2}$.}
  \label{fig:Fmax_flatdist}
 \end{center}
\end{figure}

\begin{figure}
 \begin{center}
  \includegraphics[width=8.5cm]{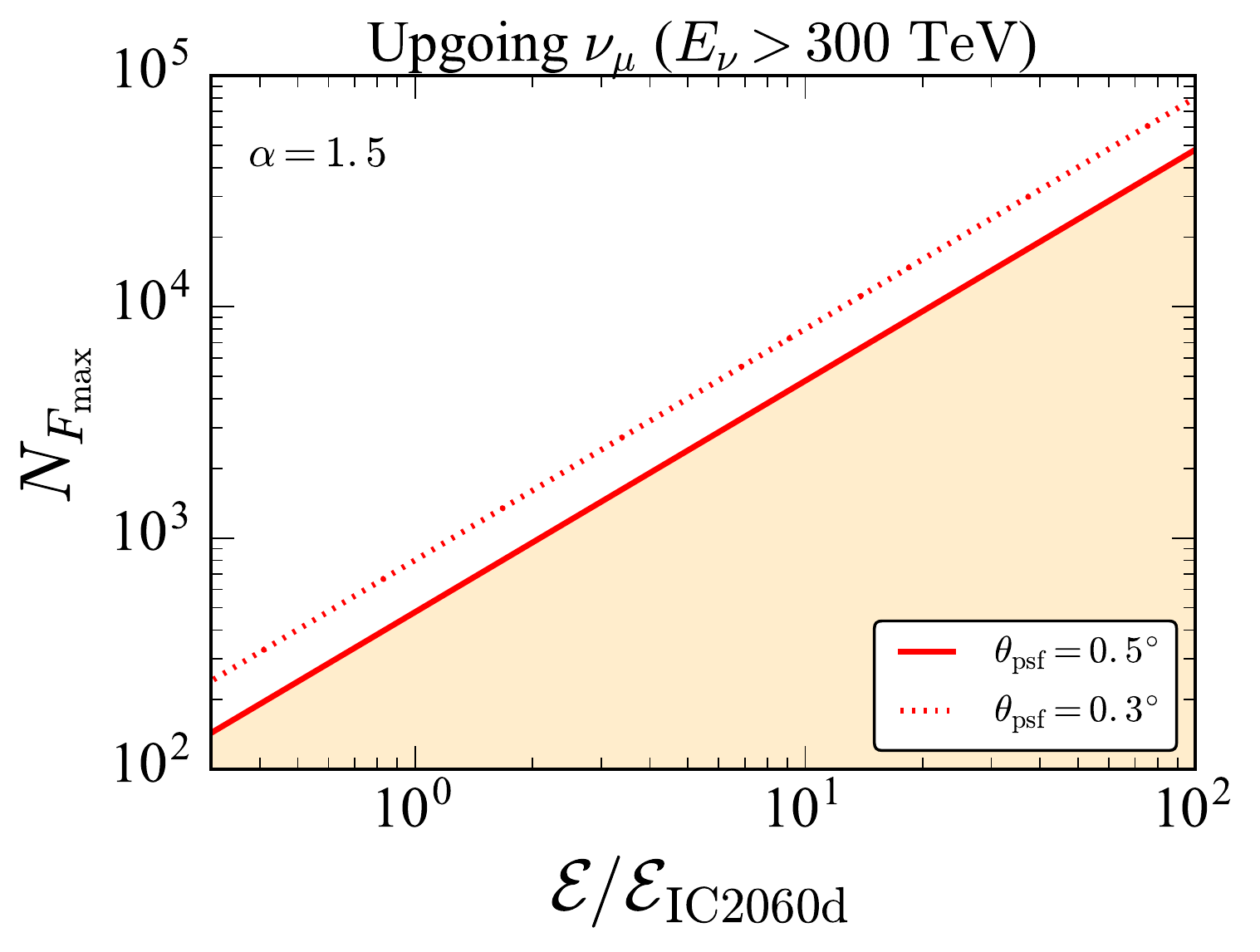}
  \includegraphics[width=8.5cm]{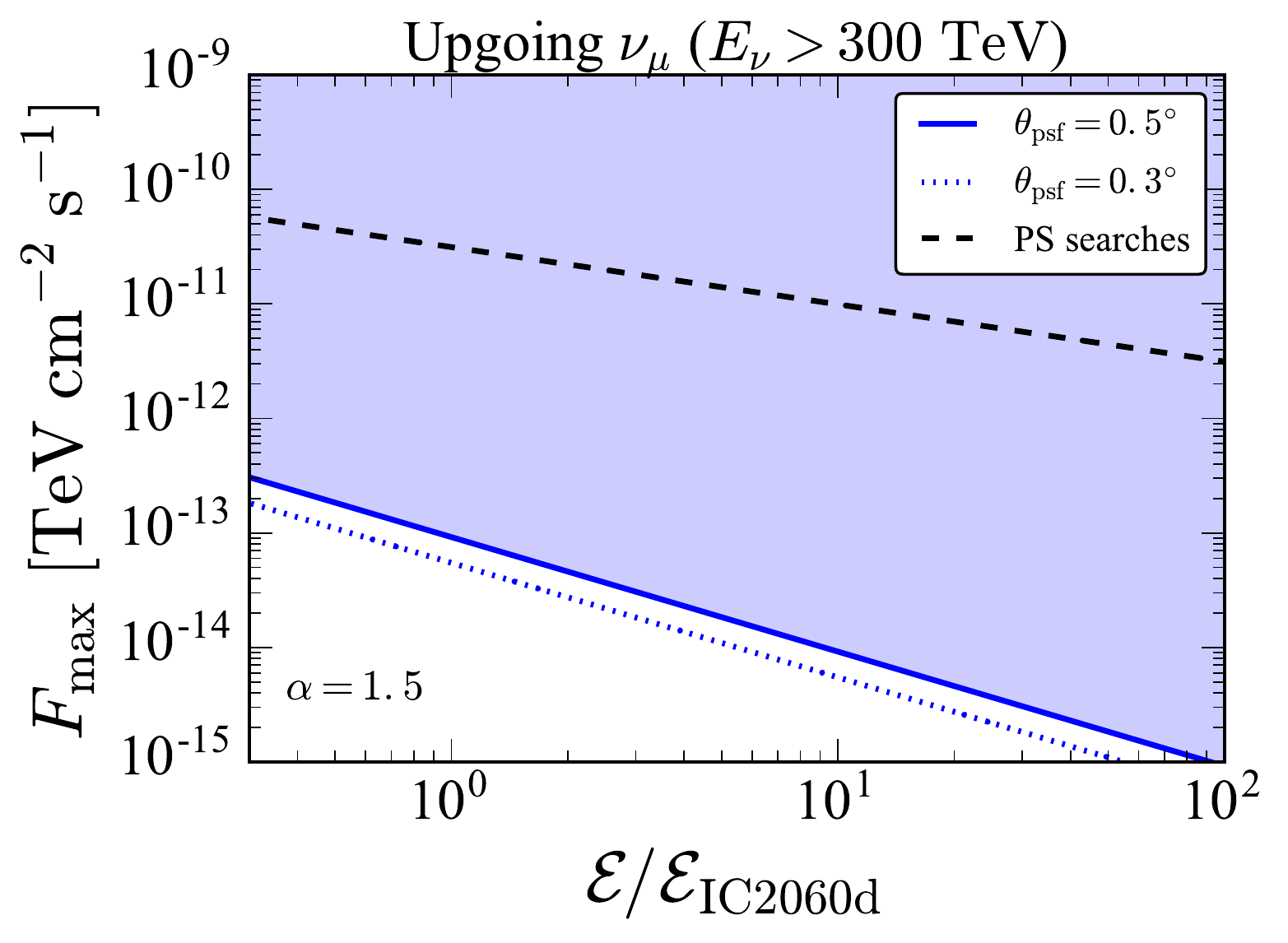}
  \caption{The same as Fig.~\ref{fig:Fmax_flatdist} but for upgoing
  $\nu_\mu$ tracks above 300~TeV, for which $\theta_{\rm psf} =
  0.5^{\circ}$ (solid) and $0.3^{\circ}$ (dotted).}
  \label{fig:Fmax_flatdist_munu}
 \end{center}
\end{figure}

In such a case of flat luminosity function exclusively at cosmological
distances ($z \agt 1$), we therefore need to re-derive the relevant
equations (\ref{eq:nuintensity}) and (\ref{eq:Poisson angular power
spectrum}), as they are based on the assumption of $2 < \alpha < 3$.
A schematic representation of the main contributions to the
distribution's first moments is shown in Fig.~\ref{fig:dNdF_flatdist}.
We find that this time, the contribution to both $I_\nu$
and $C_\nu^{P}$ is dominated by sources around $F_{\rm max}$, and hence,
by studying them, we can constrain the flux of the brightest source
$F_{\rm max}$ together.
On the other hand, $F_\ast$ would be entirely unconstrained, even if
such a break existed.
Also, since the mean intensity is dominated by $N_{F_{\rm max}}$
sources, we do not have to discuss the one-source limit, $F_{\rm
max}^{\rm 1s}$.
Corresponding to Eqs.~(\ref{eq:nuintensity}) and (\ref{eq:Poisson
angular power spectrum}), we have, for $\alpha < 2$,
\begin{eqnarray}
 I_\nu &=& \eta_3 \mathcal N_{F_{\rm max}} F_{\rm max},\\
 C_\nu^P &=& \eta_2 \mathcal N_{F_{\rm max}} F_{\rm max}^2,
\end{eqnarray}
where $\eta_3 = (2-\alpha)^{-1}$, and $\mathcal N_{F_{\rm max}} \equiv 
N_{F_{\rm max}}/(4\pi) \equiv \mathcal N_s F_{\rm max} P_1(F_{\rm
max})$.
Constraints on $F_{\rm max}$ and $\mathcal N_{F_{\rm max}}$ are then
obtained by solving these equations, given measured $I_\nu$ and upper
limit $C_{\nu,{\rm lim}}^P$:
\begin{eqnarray}
 F_{\rm max} &<& \frac{\eta_3 C_{\nu,{\rm lim}}^P}{\eta_2 I_\nu},
  \label{eq:Fmax_flatdist} \\
 \mathcal N_{F_{\rm max}} &>& \frac{\eta_2 I_\nu^2}{\eta_3^2 C_{\nu,{\rm
  lim}}^P}.
\end{eqnarray}
Again, if a fraction $k$ of the total intensity measured is attributed
to this source population, then $I_\nu$ should be replaced with $k
I_\nu$ in the equations above.

Figure~\ref{fig:Fmax_flatdist} shows the constraints on $F_{\rm max}$
and $N_{F_{\rm max}}$ as a function of exposure normalized to that of
the 4 years of IceCube operation, for $\alpha = 1.5$ and for HESE events.
Rescaling to other values of $\alpha$ is trivial by looking at
Eq.~(\ref{eq:Fmax_flatdist}); for $\alpha = 1.1$
and 1.8,
we obtain 0.7 and 2 times larger limits on $F_{\rm max}$, respectively.
Figure~\ref{fig:Fmax_flatdist_munu} is the same as
Fig.~\ref{fig:Fmax_flatdist} but for the high-energy upgoing tracks
considered in Sec.~\ref{sec:numu}, where the exposure is normalized to
the current IceCube value with the livetime of 2060 days.

The anisotropy constraints in the case of $\alpha<2$ show that the
IceCube neutrinos have to be made by at least tens to hundreds of
sources around $F_{\rm max}$.
The current upper limit on $F_{\rm max}$ from the angular power spectrum
already exceeds the point-source limit.
We note that this approach is closely related to a stacking analysis
assuming that multiple sources have the same flux, as performed in
Ref.~\cite{IceCubePS}.
Since the power spectrum is the variance, its sensitivity and hence that
to $F_{\rm max}$ improves linearly with the exposure, while that
from the point-source searches goes only as square root of the
exposure.
This makes the angular power spectrum even more important for the
next generation of neutrino telescopes.

\bibliographystyle{utphys}
\bibliography{refs}

\end{document}